\documentclass[preprint,preprintnumbers]{revtex4}

\usepackage{graphicx}

\begin{document}



\title{The  Marginal Fermi Liquid - An Exact Derivation
Based on Dirac's First Class Constraints Method }

\author{D.Schmeltzer}
\affiliation{Department of Physics,CCNY of CUNY, NY,NY,10031}
\date{\today}

\begin{abstract}
 Dirac's method for  constraints is used for solving  the problem of exclusion of double occupancy  for Correlated Electrons. The constraints are  enforced by   the pair operator $Q(\vec{x})=\psi_{\downarrow}(\vec{x})\psi_{\uparrow}(\vec{x})$  which  annihilates  the ground state $|\Psi^0>$. Away from half fillings the operator $Q(\vec{x})$  is   replaced by a set of $first$ $class$   Non-Abelian constraints  $Q^{(-)}_{\alpha}(\vec{x})$ restricted to negative energies. The propagator for a single hole  away from half fillings is determined by modified measure which is a function of the  time duration of the hole propagator.  As a result: a) The imaginary part of the self energy - is linear in the frequency.  At large hole concentrations  a Fermi Liquid self energy  is obtained. b) For the Superconducting state  the constraints generate an asymmetric  spectrum excitations between electrons and holes giving rise to   an asymmetry tunneling density of states.



Referee Comments:
\textbf{"`It is indeed refreshing to see an attempt at a completely novel route to some of these problems.  The new approach presented in the manuscript has the potential of stimulating significant further developments by other researchers. I am looking forward for others to follow in the footsteps of the ideas presented in this paper"'}

\end{abstract}


\maketitle

\section{INTRODUCTION}

The central problem in  high $T_{c}$ Superconductivity  is to treat correctly the effects of strong electron-electron interactions. We consider the zero temperature region away from half fillings in the absence of the magnetic order.  The physics in this regime is governed by  the absence   of double occupied  sites. Based on experimental results we know that once the exchange interaction is added it will generate a superconducting ground state.
 
 For a lattice model the effects of interactions are described within a repulsive Hubbard $U$ interaction. Due to the large on-site  repulsion the  double occupied state are prohibited. This means that we can project out from the electronic  spectrum the double occupied states. As a result the anti-commutation  rules  for the Fermionic operators are modified and calculations become difficult.

A significant simplification takes place in one space dimension where the method of Bosonization  shows  that for any finite Hubbard U away from half  fillings  the physics is governed by two Luttinger Liquids (one for charge and the second one  for spin).
The  limit of $U\rightarrow \infty$ can not  be considered in a  microscopic formulation. This limit can be taken for the renormalized model within  the Renormalization  Group ($R.G.$)  calculations. One obtains  a line  of  Luttinger fixed points.

 The wave function has been obtained  from the   Bethe ansatz  \cite{Shiba} in the limit $U/t\rightarrow \infty$ (where $t$ is the hopping constant).  This solution shows  that the spin configuration becomes degenerate at $U=\infty$ and the wave function is a $singlet$ ground state for all values of  $0<U<\infty$. This means that the degeneracy at $ U=\infty$ is removed by any infinitesimal  perturbation $t/U$.
A formal way for describing the wave function  is to say that the ground state is annihilated by the singlet operator $Q(\vec{x})|\Psi^{0}>=0$, $Q(\vec{x})\equiv\mathbf{\psi}_{\uparrow}(\vec{x})\mathbf{\psi}_{\downarrow}(\vec{x})$.  The solution of this  last equation shows that the wave function can be written as a product of a singlet state with  another unknown state.
 
For two space dimensions we do not have an exact solution as a function of the Hubbard $U $. Therefore we have to consider  the projection of double occupancy which forces us to deal with the question of the modified commutators.
A direct approach for dealing with the modified commutators is given by  the Hubbard  $X$ operators. These operators are a mixture of $Bosonic$ and $Fermionic$ excitations.  As expected, this leads  to a complicated representation which is  difficult to handle \cite{Zai}.
The conventional wisdom in higher dimensions is the   method of  the slave particles  \cite{Barnes} (slave Fermions, or slave Bosons). The slave particles representation replaces the exclusion of double occupancy by two slave fields  (one for charge and one for spin). These excitations are coupled  by a  $U(1)$ gauge field  \cite{David,Wiegmann, Baskaran,Lee}. For space  dimensions $d\geq 2$   the gauge field is in the $confined$ phase, causing  the slave particles  to  be   strongly coupled. For  some special conditions a $deconfined$ phase might   be possible \cite{FSch,Fisher} .
The slave particles representation works in one dimension  \cite{Sch} since the guage field is in the $deconfined$ phase.  In this phase, the excitations are described by solitons which carry  fractional quantum numbers \cite{Nayak}.
An explicit solution based on  the slave-boson method for the Hubbard  $U=\infty$ case has been considered  in the literature  by  \cite{Sch} within a path integral formulation.
This formulation has been criticized in   \cite{Kopp} which  argued that  the  $path$ $integral$ $measure$  for the slave particles  is incorrect.
The price   we pay when we work with the  slave particles  is that the single particle   spectrum   is  described by a pair of $non$  $physical$ $excitations$!
  
An alternative approach  for dealing with the large $U$ repulsion interaction  is to use  the Gutzwiller  projection method used by    \cite {Gutz,Muthu}. Using this  method  combined  with a variational procedure, the authors  \cite {Muthu} have constructed a  variational  wave function for   the strongly correlated superconductors. 
It has been pointed out  \cite{Anderson} that the exclusion of double occupancy is responsible  for the strong asymmetry between the hole and the electronic  excitations observed in the  tunneling spectrum for the optimally doped BSCCO. 

One of the successful  phenomenological theories used to explain a varieties of experiments  is the marginal Fermi liquid theory \cite{Varma,VarmAji},  yet the relation of this model to the microscopic theory  is  not clear.

 The recent  quantum oscillations observed in the Shubnikov de  Haas 
experiment might raise questions about the validity of different approaches and in particular it might test the validity  of the projected wave function  \cite{Doiron,Julian}.

 The purpose of the present  paper is to introduce a new method for dealing with the problem of exclusion double occupancy.
We propose to use Dirac's theory for $First$ $Class$ $constraints$ \cite{Dirac}.
The solution of the problem will be formulated in the language of Quantum constraints : 
for a hamiltonian $H$ and a constraint operator  $Q(\vec{x})=\mathbf{\psi}_{\uparrow}(\vec{x})\mathbf{\psi}_{\downarrow}(\vec{x})$ one has to find the many body state  $|\Psi^{0}>$ which satisfies,
$H|\Psi^{0}>=E|\Psi^{0}> $ and  is annihilated  by the constraint operator $Q(\vec{x})|\Psi^{0}>=0$.
Since  the constraint should be satisfied at any time   the time derivative of the operator   $Q(\vec{x})$   requires that    the commutator $[Q(\vec{x}),H] $  must  vanish. Away from half fillings we restrict the constraints to negative energies (holes  excitations) and obtain a set of $first$ $class$  non-Abelian  constraints   $ Q^{(-)}_{\alpha}(\vec{x})$ $\alpha=1,2,3$  which obey  $Q^{(-)}_{\alpha}(\vec{x})|\Psi^{0}>=0$. 
In order to have a canonical theory in the presence of the constraints, we enlarge the Hilbert space by including new anti-commuting fields  \cite{Teitelboim,Holten}. As a result when a hole is created at a time $t_{i}$ and destroyed at the time $t_{f}$  the evolution in the enlarged space is canonical. The Physical processes  occur at  times  $t_{i}$ and  $t_{f}$ in the Physical Hilbert space. 
The projection into the physical  Hilbert space is done by using proper boundary conditions for the anti-commuting fields  \cite{Teitelboim,Holten}.  Due to the non-commutativity of the constraint, the projection will generate a time dependent non-linear measure  for the Lagrange multipliers. 
As a result, the physical evolution operator  for the hamiltonian $H$ with constraints   is given by,
$\hat{U}_{phys}[t_{f},t_{i}]=e^{\frac{-i}{\hbar}(t_{f}-t_{i})H
\int \prod_{\alpha=1}^{3}\mathcal{D}\lambda_{\alpha}(T) e^{\frac{-i}{\hbar}}\sum_{\alpha=1}^{3}\int\lambda_{\alpha}(\vec{x})Q^{(-)}_{\alpha}(\vec{x})\,d^{d}x}$

where $\prod_{\alpha=1}^{3}\mathcal{D}\lambda_{\alpha}$ is a $non-linear$ $measure $ which is generated by projection into the Physical Hilbert space for  the non-commuting constraints.   The
effective interaction for the holes   sectors  depend explicitly on the time interval  
$T=t_{i}-t_{f}$. 

 The effective non local action obtained from the temporal projection is investigated  with the help  of the R.G. method. We find that the single hole excitation has  a width which is  linear in frequency and  the  scattering rate  obeys   $\frac{1}{\tau}\propto \omega F(\frac{\omega}{v_{F}\Lambda})$ in agreement with  the infrared data \cite{Schlesinger}. The linear frequency  width is controlled by the holes density. When  the hole density increases the region of the linear frequency width shrinks;   for large densities it shrinks to zero and therefore a Fermi liquid behavior is obtained.  
The theory presented in this paper is applicable away not at half fillings  for densities $x>x_{c}$ at   zero temperature where the magnetic order has been  suppressed. Therefore we will not investigate the metal insulator transition. For a finite hole concentration the addition of a finite exchange interaction   $j\neq 0$  (in the $t-j$ model) will give rise to a superconducting phase \cite{hdavid}. An important   question will be to understand the effect of exclusion of double occupancy on the superconducting state.
We will show that  by projecting out double occupancy, an asymmetry of tunneling density of states is observed.
Using this theory we will explain  the asymmetric tunneling density of states observed  by   \cite{Pan,Ando}.

The content  of this paper is as following:
In chapter II we present the adaptation of  the method of First Class constraints   \cite{Dirac,Vilkov,Batalin,Teitelboim,Holten,Fulp} to Condensed Matter Physics.   
In section III we present the model for correlated electron and show that the ground state can be obtained within the method of quantum constraints.  The exclusion  of the double occupied states has been investigated  in the literature  \cite{Gutz,Muthu} using  the Gutzviller  projection method.  The solution of the  equation $Q(\vec{x})|\Psi^{0}>=0$ is described by the  $Jastrow$ representation    \cite{Gutz,Muthu}. In chapter  IV we show that the method of first class    allows for additional    non-Abelian constraints which  give rise to a non-linear  integration measure and  replaces the  delta function constraint used implicitly  in refs.\cite{Gutz,Muthu}.  Using the first class constraint  method we  compute the ground state for  the high $T_{c}$ superconductors. Chapter V is devoted to the computation of the integration measure. In chapter VI we introduce the canonical phase space action for the $t-j$ model. In chapter VII we consider the explicit case where the exchange interaction is zero.  In chapter  VIII we introduce two  Green's functions: $G$  (the physical one) and $D$ (the parametrical one). Due to the projection into the physical Hilbert space, the action which governs the single hole propagation is time dependent. The physical Green's function $G$ is computed in terms of a parametric Green's function $D$. 
In chapter $IX$ we perform the R.G. calculation  for the effective action at  the  time  interval  $T=t_{i}-t_{f}$ using the finite size scaling.   Chapter $X$ is devoted to the calculation of the parametric Green's function $D$.
In chapter $XI$ we present the calculation for the physical Green's function $G$. In particular we show that the relaxation rate is linear in frequency. We believe that this  explains the experimental results observed for the optical conductivity  given in 
\cite{Schlesinger}.
Finally in chapter $XII$  we consider the effect of the projected Green's function on   the  superconductor ground state . We add a $pairing$ interaction  and compute the tunneling density of states for a superconductor using the projected Green's function. We show that the projection  (incorporated into the calculation  trough the self energy) gives rise to an  asymmetric tunneling   density of states, in agreement with the experiments \cite{Pan,Ando}.

In chapter $XIII$ we present our conclusion. We have included an  Appendix where we  discuss  the effects of the secondary first class constraints generated by the commutators between the hamiltonian and the primary constraints.

\vspace{0.3 in }

\section{THE  METHOD OF QUANTUM   CONSTRAINTS} 

\vspace{0.3 in }

The purpose of this  chapter is to present an adaptation of the method of quantum  constraints to Condensed Matter Physics.

  We have   to find the ground state  $|\Psi^{0}>$ under the conditions that a set of operators $Q_{\alpha}(\vec{x})$,  $\alpha=1,2....n$    are restricted to be zero!
In Quantum Mechanics   this means that  one has to find   the  state  $|\Psi^{0}>$  which is annihilated by the constraints $Q_{\alpha}(\vec{x})|\Psi^{0}>=0$ and is an eigenstate of the hamiltonian $H|\Psi^{0}>=E|\Psi^{0}>$.

 There  are two types of  constraints : 

\textbf{Second Class  constraints } \cite{Dirac}  are characterized by  a non singular matrix   $[Q_{\alpha}(\vec{x}),Q_{\alpha'}(\vec{x})]$ where the symbol    $[,]$ represents  the commutator  for the Bosonic constraints. The matrix $[Q_{\alpha}(\vec{x}),Q_{\alpha'}(\vec{x})]$ can be inverted and therefore  has  a non vanishing determinant    $det[Q_{\alpha}(\vec{x}),Q_{\alpha'}(\vec{x})]\neq 0$ for $\alpha,\alpha'=1,...m$  and  $m=\frac{n}{2}$. 
Recently we  have used   this method to compute the  persistent currents in coupled rings \cite{genus} and study  mesoscopic vortices  in a two dimensional electron gas \cite{PMC}.

 When the determinant of the constraints vanishes we obtain \textbf{First Class  constraints} which will be used  to solve the problem of exclusion of double occupancy. 
 According to  Dirac  \cite{Dirac} one has to identify all the constraints which must be satisfied at   any time.   
\begin{equation}
\frac{d}{d t}Q_{\alpha}(\vec{x},t)|\Psi^{0}>=0
\label{eqconstrainned}
\end{equation}

From  the Heisenberg equation of motion  we obtain,
$\frac{d}{d t}Q_{\alpha}(\vec{x},t)=\frac{1}{i \hbar}[Q_{\alpha}(\vec{x},t),H]$ where the new constraints are given by the difference between the commutator and a linear combination of the existing constraints,
$[Q_{\alpha}(\vec{x},t),H]- \sum_{n=1}^{n} T^{\alpha}_{\beta}Q_{\beta}(\vec{x})$$=\sum_{n=1}^{r}t^{\alpha}_{\beta}q_{\beta}(\vec{x})$.  
In this equation  $T^{\alpha}_{\beta}$ and $t^{\alpha}_{\beta}$ stand  for a set of  matrix elements and $q_{\beta}(\vec{x})$, $\beta=1,2..r$ represent the new (generated)  secondary  first class constraints.
For  the remaining part we will represent  the two sets  $\alpha=1,2..n$  (the primary first class  constraints) and $\beta=1,2..r$  (secondary  first class constraints)  by one set    $(Q_{1}(\vec{x},t),Q_{2}(\vec{x},t),...Q_{n}(\vec{x},t),q_{1}(\vec{x}),q_{2}(\vec{x}),...q_{r}(\vec{x}))\Rightarrow Q_{\alpha}(\vec{x},t)$ where  $\alpha=1,2..(n+r)$  is  the new index for the two sets.  
For the remaining part we will assume that  $Q_{\alpha}(\vec{x},t)$  are $all$ the first class constraints (no new constraints are generated by higher order commutators of the hamiltonian with  all the $n+r$ constraints). 

Since the  commutator of the constraints can  be zero,  the  inverse of the commutator does not exists. As a result, a modification of the commutation rules  as is  done for Second Class constraints  is not possible \cite{Dirac}.

To overcome this difficulty   one introduces       $new$  $constraints$  $\Phi_{\beta}(\vec{x},t)$, $\beta=1,2,..(n+r)$, 
\begin{equation}
\Phi_{\beta}(\vec{x},t)|\Psi^{0}>=0  
\label{phi}
\end{equation}

In order to obtain a  canonical phase space \cite{Itz1} with less variables, we have to project out  an even number of constraints (the constraints $Q_{\alpha}(\vec{x})$ and their canonical  conjugate one $\Phi_{\alpha}(\vec{x},t)$). 
This is achieved   by demanding that the  determinant of the  commutator in the enlarged  Hilbert space (with the additional unknown   constraints $\Phi_{\alpha}(\vec{x},t)$) is not zero.

\begin{equation}
Det[Q_{\alpha},\Phi_{\beta}]\neq 0 
\label{det}
\end{equation}

As a result we obtain an equivalent  theory  with a fewer  $independent$ degrees of freedom \cite{Itz1}.
Using this conditions we can  compute the \textbf{The Quantum Evolution Operator}.

a) The  Quantum Evolution Operator for the  $unconstrained$ case is given by 
 $\hat{U}[t_{f},t_{i}]=e^{\frac{-i}{\hbar}(t_{f}-t_{i})H}$.

The matrix elements of the evolution  operator are  computed according to the path integral method for  Grassmann anti-commuting functions  \cite{Itz2}.
 Using the Grassmann Coherent states one introduces states $|\psi_{\sigma}(\vec{x})>$
 ,$<\overline{\psi}_{\sigma}(\vec{x})|$ which obey; $\mathbf{\psi}_{\sigma}(\vec{x})|\psi_{\sigma}(\vec{x})>=\psi_{\sigma}(\vec{x})|\psi_{\sigma}(\vec{x})>$  and    
$<\overline{\psi}_{\sigma}(\vec{x})| \mathbf{\psi}^{\dagger}_{\sigma}(\vec{x})=<\overline{\psi}_{\sigma}(\vec{x})|\overline{\psi}_{\sigma}(\vec{x})$ where 
 $(\overline{\psi}_{\sigma}(\vec{x}))^2=(\psi_{\sigma}(\vec{x}))^2=0$.
 
This allows us  to formulate a field theory in the  $Schroedinger$ $representation$ where the role of the $coordinate $ is played by $\psi_{\sigma}(\vec{x})$   and the $canonical$ $conjugate$  momentum  $\mathbf{\psi}^{\dagger}_{\sigma}(\vec{x})$ is given by 
  $\mathbf{\psi}^{\dagger}_{\sigma}(\vec{x})=\frac{\delta}{\delta\psi_{\sigma}(\vec{x})}$.

 The matrix elements of the  quantum evolution  operator in the Grassmann space are given by
 $<\overline{\psi},t_{f}|\psi,t_{i}>=< \overline{\psi}|\hat{U}[t_{f},t_{i}]|\psi>\equiv U[\overline{\psi}_{f},t_{f};\psi_{i},t_{i}]$.

Following  \cite{Itz2} we obtain the path integral  representation for  the matrix element 
$U[\overline{\psi}_{f},t_{f};\psi_{i},t_{i}]$:
\begin{eqnarray}
U[\overline{\psi}_{f},t_{f};\psi_{i},t_{i}]&=&\int D(\overline{\psi}_{\sigma},\psi_{\sigma}) e^{(\frac{1}{2}(\overline{\psi}_{f}\psi_{f}+\overline{\psi}_{i}\psi_{i})+i\int_{t_{i}}^{t_{f}}[\frac{1}{2i}(\overline{\psi}_{\sigma}\frac{d\psi_{\sigma}}{\,dt}-\frac{d\overline{\psi}_{\sigma}}{\,dt}\psi_{\sigma})-H(\overline{\psi}_{\sigma},\psi_{\sigma},t)]\,dt)}
\end{eqnarray}

b) The Quantum Evolution Operator for the $constrained$  system will be given  in terms of  evolution matrix elements  $U_{phys}[\overline{\psi}_{f},t_{f};\psi_{i},t_{i}]$ in the Grassmann space. Due to the  new constraints  Faddeev  \cite{Fad} has shown that  a $Superdeterminant$
is needed for the integration measure. Such a formalism has been used by \cite{Greco}.

 A simpler method   is to represent the   physical  evolution operator  in terms of $only$ the physical constrained $Q_{\alpha}$. The integration with respect to  the $unknown$  $\Phi_{\beta}$ constraints     will  modify the integration measure from  
 $d \lambda_{\alpha}$  to  a $non-linear$ integration  measure $\mathcal{D}\lambda_{\alpha}(T=t_{i}-t_{f})$. 
 This  allows  to represent the $physical$ $evolution$ operator in a  form   which is similar with to the result given  by \cite{Klauder}:

\begin{equation}
\hat{U}_{phys}[t_{f},t_{i}]=e^{\frac{-i}{\hbar}(t_{f}-t_{i})H}
\int \prod_{\alpha=1}^{(n+r)}\mathcal{D}\lambda_{\alpha}(T) e^{\frac{-i}{{\hbar}}\sum_{\alpha=1}^{(n+r)}\int\lambda_{\alpha}(\vec{x})Q_{\alpha}(\vec{x})\,d^{d}x}
\label{final}
\end{equation}

Eq.(5) emerges from the canonical phase space formalism  which we will present in the remaining part of this chapter.

We introduce the  Lagrange multipliers, $\lambda^{\alpha}(\vec{x})$,  $\alpha=1,2,..(n+r)$ to enforce the exclusion of double occupancy.  In addition we seek new constraints which are canonical conjugates to the original ones.
The new  constraints $\Phi_{\beta}(\vec{x},t)$, $\beta=1,2,..(n+r)$ are introduced with the help of new  Lagrange multipliers $\pi_{\alpha}(\vec{x})$ ,$\alpha=1,2,..(n+r)$.
The $complete$  Hamiltonian which contains both type of  constraints is given by the  hamiltonian $H_{T}$:
  
\begin{equation}
H_{T}=H+\int[\sum _{\alpha=1}^{(n+r)}\lambda^{\alpha}(\vec{x})Q_{\alpha}(\vec{x})
 + \sum _{\alpha=1}^{(n+r)}\pi_{\alpha}(\vec{x})\Phi^{\alpha}(\vec{x})]\,d^{d}x
\label{ total hamiltonian}
\end{equation}

It is convenient to  replace the constraint $\Phi^{\alpha}(\vec{x})$ by an $equivalent$  constraint   $\chi^{\alpha}(\vec{x})$:
\begin{equation}
\Phi^{\alpha}(\vec{x},t)=-\frac{d}{dt}\lambda^{\alpha}(\vec{x},t)+\chi^{\alpha}(\vec{x},t)
\label{change}
\end{equation}

As result,  the transformed  Hamiltonian  which contains  the time derivative of the Lagrange multiplier $\lambda^{\alpha}(\vec{x})$  will be modified.

Using the results given in equation $(6)$ and $(7)$ we obtain a $new$  formulation of the constraint problem.
At this point it is preferable  to work with the  canonical phase space momentum-coordinate action $S$.  $S=\int \,d^{d}x\int\,dt L$  where $L$ is the Lagrangian and  $h(\vec{x})$ is  the hamiltonian density for the hamiltonian $H$. 
\begin{equation} 
L=\sum_{\sigma=\uparrow,\downarrow}i\hbar\psi^{\dagger}_{\sigma}(\vec{x},t)\partial_{t}\psi_{\sigma}(\vec{x},t)+\sum_{\alpha=1}^{(n+r)}\pi_{\alpha}(\vec{x},t)\partial_{t}\lambda^{\alpha}(\vec{x},t)-h(\vec{x},t)-\sum_{\alpha=1}^{(n+r)}(\pi_{\alpha}(\vec{x},t)\chi^{\alpha}(\vec{x},t)+\lambda^{\alpha}(\vec{x},t)Q_{\alpha}(\vec{x},t))
\label{action}
\end{equation} 
 The term  $\pi_{\alpha}(\vec{x},t)\partial_{t}\lambda^{\alpha}(\vec{x},t)$ allows to identify  the canonical conjugate momentum of $\lambda^{\alpha}(\vec{x})$ with $\pi_{\alpha}(\vec{x})$ which obeys the commutation rules,  $[\lambda^{\alpha}(\vec{x}),\pi_{\beta}(\vec{y})]=i\hbar\delta_{\alpha,\beta}\delta(\vec{x}-\vec{y})$.

It is important to point out  that equation $(8)$ can be understood as a starting point for our theory where the Lagrange multiplier  $\lambda^{\alpha}(\vec{x})$ enforces the constraints  $Q_{\alpha}(\vec{x})$. The Quantum nature of the Lagrange multipliers is enforced by demanding the existence of a canonical conjugate variable  $\pi_{\alpha}(\vec{x})$. Physically  the canonical conjugate momentum must be enforced to be zero. This is done by introducing a new Lagrange multiplier $\chi^{\alpha}(\vec{x})$.
As a result our theory will have the set of constraint equations:

$\pi_{\alpha}(\vec{x})|\Psi^{0}>=0$ and  $Q_{\alpha}(\vec{x})|\Psi^{0}>=0$.

Which will be enforced by the two sets of Lagrange multipliers:
 $\chi^{\alpha}(\vec{x})$  and  $\lambda^{\alpha}(\vec{x})$.

 The new  Lagrange multipliers   $\chi^{\alpha}(\vec{x})$, $\alpha=1,2,..(n+r)$ have not yet  been specified.  
$ VILKOVISKY$ \cite{Vilkov}  has proved  a  theorem  which shows that the role of the new Lagrange multipliers  $\chi^{\alpha}(\vec{x})$  is equivalent  to the gauge  fixing in Quantum Electrodynamics. 
The proof of the theorem \cite{Vilkov, Batalin} is based on the followings steps:
 
1. The action given in eq.$(8)$ can be represented  in an $equivalent$ way using new fermionic   fields.  One introduces a pair of anti-commuting fields  for each one of the constraints $\pi_{\alpha}(\vec{x})|\Psi^{0}>=0$ and  $Q_{\alpha}(\vec{x})|\Psi^{0}>=0$.

2. The  Lagrange multiplier $\chi^{\alpha}(\vec{x})$  plays the role of gauge  fixing condition. The replacement of the action in equation  $( 8)$ with a new action written in terms of the new   Fermionic  fields show  explicitly that the expectation values for any physical observable is invariant under a change of the  Lagrange multipliers $\chi^{\alpha}(\vec{x}) \rightarrow \chi^{\alpha}(\vec{x}) +\delta (\chi^{\alpha}(\vec{x}))$.  Therefore the physical results are $independent$ of  the particular choice of the fields $\chi^{\alpha}(\vec{x})$.

\textbf{An  exact mapping  of the action given in  equation $(8)$  to an equivalent  action where the constraints are replaced   by  new  Fermionic fields is  possible \cite {Vilkov,Batalin,Holten}.   As a result one obtains  an enlarged Hilbert space of      anti-commuting   fields.} 

We consider the case  where  the constraints  $Q_{\alpha}(\vec{x})$ obey the following relations:
\begin{equation}
[Q_{\alpha},Q_{\beta}]=F^{\gamma}_{\alpha,\beta}(Q_{\gamma})
\label{nlc}
\end{equation}
Where $F^{\gamma}_{\alpha,\beta}(Q_{\gamma})$ is a nonlinear function of the constraint $Q_{\gamma}$  such
that  for $ Q_{\gamma}=0$ we have the result $F^{\gamma}_{\alpha,\beta}(Q_{\gamma}=0)=0$ for any $\alpha$, $\beta$ and $\gamma$.
 
For this case the  $ VILKOVISKY$ \cite{Vilkov} theorem  allows the exact mapping of the action in eq.$(8)$  to the new action defined in terms of the  new anti-commuting  fields.

The mapping is done according to the following steps  \cite{Holten}:

\vspace{0.1 in}

a) For each constraint field  $Q_{\alpha}(\vec{x})$ one introduces a pair of anti- commuting real fermions $C^{\alpha}(\vec{x})$ and $b_{\alpha}(\vec{x})$  

\begin{equation}
[C^{\alpha}(\vec{x}),b_{\beta}(\vec{x'})]_{+}=\delta_{\alpha,\beta}\delta(\vec{x}-\vec{x'})
\label{fghost}
\end{equation}
Such that $b_{\beta}(\vec{x'})$ acts as an annihilation operator. 
\vspace{0.1 in}

b) Similarly  the  canonical momentum constraints  $\pi_{\alpha}(\vec{x})$ is replaced   by  the pair of anti-commuting real fermions $e^{\alpha}(\vec{x})$ and $f_{\alpha}(\vec{x})$.  

\begin{equation}
[e^{\alpha}(\vec{x}),f_{\beta}(\vec{x'})]_{+}=\delta_{\alpha,\beta}\delta(\vec{x}-\vec{x'})
\label{ehhost}
\end{equation}
\vspace{0.1 in}
c) \textbf{The physical Hilbert space is extended  to an enlarged Hilbert space.} Therefore the  many body state $|\Psi ^{0}\rangle $ is replaced by a new  
state $|\Psi> $  which  is build from   a Fermionic subspace needed to enforce  the constraints. The wave function $<\psi_{\sigma}(\vec{x})|\Psi^{0}>$ is replaced  by  a wave function in the $enlarged$ Hilbert space 

\begin {equation}
<\psi_{\sigma}(\vec{x}),C^{\alpha}(\vec{x});\pi_{\alpha}(\vec{x}), e^{\alpha}(\vec{x})|\Psi>
\label{waf}
\end{equation}

Where $<\psi_{\sigma}(\vec{x}),C^{\alpha}(\vec{x});\pi_{\alpha}(\vec{x}), e^{\alpha}(\vec{x})|$  is the $coherent$ $state$ representation in the enlarged Hilbert space.

\vspace{0.1 in}

d) According to the theorem (see pages 247, 322-324 in \cite{Teitelboim}) the physical wave function $<\psi_{\sigma}(\vec{x})|\Psi^{0}>  $ is obtained by  the  projection 
$<\psi_{\sigma}(\vec{x}),C^{\alpha}(\vec{x})=0;\pi_{\alpha}(\vec{x})=0, e^{\alpha}(\vec{x})=0|\Psi>$. 

\textbf{The  physical evolution operator matrix elements     $U_{phys}[\overline{\psi}_{f},t_{f};\psi_{i},t_{i}]$ are  obtained  once we impose the temporal boundary conditions on the auxiliary fields in  the  $enlarged$ Hilbert space.} We use the following $temporal$ boundary conditions:

\begin{equation}
C^{\alpha}(\vec{x},t_{i})=C^{\alpha}(\vec{x},t_{f})=0 
\label{ctictf}
\end{equation}

\begin{equation}
e^{\alpha}(\vec{x},t_{i})=e^{\alpha}(\vec{x},t_{f})=0 
\label{eief}
\end{equation}

For the momentum $\pi_{\alpha}(\vec{x},t)$ we use  the following  boundary conditions:

\begin{equation}
\pi_{\alpha}(\vec{x},t_{i})=\pi_{\alpha}(\vec{x},t_{f})=0 
\label{pip}
\end{equation}

\vspace{0.1 in}

e) In the extended Hilbert space  the wave function $|\Psi^{0}>$  is  replaced  by $ |\Psi>$. 

\textbf{In the enlarged Hilbert space we replace the constraints  $Q_{\alpha}(\vec{x})$ and $\pi_{\alpha}(\vec{x})$  by a new constraint  $\Omega$.}

The constraint $\Omega$ operator has to obey

\begin{equation}
\Omega(\vec{x}) |\Psi>=0
\label{cQ}
\end{equation}

\vspace{0.1 in}

f) In the extended Hilbert space the hamiltonian $H$ is replaced by $H_{c}$. The  operator $\Omega$ obeys the extended Heisenberg equation of motion:

\begin{equation}
i\hbar\frac{d}{dt}\Omega|\Psi>=[\Omega,H_{c}]|\Psi>=0
\label{hamiltonianC}
\end{equation}
 
The extended constrained operator $\Omega$ must be NILPOTENT

\begin{equation}
\Omega^{2}=[\Omega,\Omega]_{+}=0
\label{nilpotency}
\end{equation}

\vspace{0.1 in}

g) In order to satisfy the NILPOTENCY condition the operator   $\Omega$ is  written as a sum of two parts 

\begin{equation}
\Omega =\Omega _{0}+\Omega
_{NL}
\label{ccc}
\end{equation}

Where $\Omega _{0}$ is given by

\begin{equation}
\Omega
_{0}=\sum_{\alpha=1}^{(n+r)} C^{\alpha}(x)Q_{\alpha}(x)+\sum_{\alpha=1}^{(n+r)} f^{\alpha}(x)\pi_{\alpha}(\vec{x}) 
\label{omego}
\end{equation}

\textbf{Due to the non commutativity of the constraints $[Q_{\alpha},Q_{\beta}]=F^{\gamma}_{\alpha,\beta}(Q_{\gamma})$  the 
condition $\Omega^2=0$ requires   that the constraint  operator should have a non linear part given by $\Omega_{NL}$.} 
The most general form of  the non linear part $\Omega_{NL}$ is given by: 

\vspace{0.1 in}

$\Omega_{NL}(\vec{x})=\sum_{\alpha=1}^{(n+r)} C^{\alpha}(\vec{x})[\int \,d^{d}x^{1}\sum_{\alpha_{1}=1}^{(n+r)} \sum_{\beta_{1}=1}^{(n+r)}..\int \,d^{d}x^{(n+r)}\sum_{\alpha_{n=1}}^{(n+r)} \sum_{\beta_{n=1}}^{(n+r)}\frac{i^{(n+r)}}{2(n+r)!}$
\begin{equation}
\cdot C^{\alpha_{1}}(\vec{x}^{1})..C^{\alpha_{(n+r)}}(\vec{x}^{(n+r)})M_{\alpha \alpha_{1}..\alpha_{n+r}}^{\beta_{1}..\beta_{n+r}} b_{\beta_{1}}(\vec{x}^{1})..b_{\beta_{(n+r)}}(\vec{x}^{(n+r)})]
\label{nla}
\end{equation}

\textbf{ The matrix $M_{\alpha \alpha_{1}..\alpha_{n+r}}^{\beta_{1}..\beta_{n+r}}$  elements are  determined by  the condition  $\Omega^2=0$.}

\vspace{ 0.1 in}

h) The hamiltonian  $H_{c}$  represents the extention of $H$ for the extended Hilbert space: 
\begin{equation}
H_{c}=H+i\int \,d^{d}x\sum_{\alpha=1}^{(n+r)}\sum_{\beta=1}^{(n+r)}C^{\alpha}(\vec{x})
h^{\beta}_{\alpha}(\vec{x})b_{\beta}(\vec{x})+...
\label{parameters}
\end{equation}

The parameters $h^{\beta}_{\alpha}(\vec{x})$ are determined  by  the equation  $[H_{c},\Omega]=0$.

\vspace{ 0.1 in}

k) The condition  $\Omega|\Psi>=0$   guarantees that $\Omega O_{phys}|\Psi>=[\Omega, O_{phys}]|\Psi>$.  As a result any physical operator $O_{phys}$  is mapped by   the operator $\Omega$ to another physical state,  therefore we have  $[\Omega, O_{phys}]=0$.  This implies that if $|\Psi>$ is a physical state, then 
$\Omega O_{phys}|\Psi>=[\Omega, O_{phys}]|\Psi>=0$.  

 This condition shows that any modified physical operator $O_{phys}'$ which is related to the original physical operator $O_{phys}$ trough the   transformation  
$O_{phys}'=O_{phys}+[\Omega,W]_{+}$  has the same matrix elements  as the original operator. The symbol  $[,]_{+}$ stands for the anti-commutator and $W$ is a  new  fermionic operator defined   in terms of the Lagrange multipliers $\lambda^{\alpha}(\vec{x})$ and $\chi^{\alpha}(\vec{x})$ and the  Fermionic fields.

\begin{equation}
W=\int d^{d} x\sum_{\alpha=1}^{(n+r)}[b_{\alpha}(\vec{x})\lambda^{\alpha}(\vec{x})+e_{\alpha}(\vec{x})\chi^{\alpha}(\vec{x})]
\label{ccc}
\end{equation}

The proof that the operator  $O_{phys}'=O_{phys}+[\Omega,W]_{+}$ and the operator   $O_{phys}$    have  the same matrix elements  in the extended Hilbert space  follows from the  identity  $[\Omega,[\Omega,W]]=[\Omega^2,W]=0$ (the nilpotency of $\Omega$).

\vspace{ 0.1 in}

m) This means that the  matrix elements of the hamiltonian   $H_{effective}$ are the same as  for the hamiltonian  $H_{c}$.
\begin{equation}
H_{effective}=H_{c}+[\Omega,W]_{+}\equiv H_{c}+\delta^{(+)}_{\Omega}W
\label{act}
\end{equation}

This shows that  $\delta^{(+)}_{\Omega}$ acts as  an  $exterior$ $derivative$  \cite{Teitelboim}, $\delta^{(+)}_{\Omega}A=[\Omega,A]_{+}$
where $A$ is  an arbitrary operator. 

\vspace{0.1 in}

n) Following  the  theorem \cite{Vilkov} we have the  freedom to  choose  any bosonic fields $\chi^{\alpha}$. 
In particular  the path integral is independent on the choice of the unknown Lagrange multiplier  $\chi^{\alpha}(\vec{x})$.
The path integral is invariant under the transformation $\chi^{\alpha}(\vec{x})\rightarrow (\chi^{\alpha}(\vec{x}))'=\chi^{\alpha}(\vec{x})+\delta(\chi^{\alpha}(\vec{x}))$.

As a result of this theorem one can show \cite{Teitelboim} that the quantum  expectation value  of any operator $A(t)$ which commutes with the constraint $[A,\Omega]=0 $  is independent of the choice $\chi^{\alpha}$!

\begin{equation} <A(t)>_{\chi^{\alpha}}=<A(t)>_{\chi^{\alpha}+\delta(\chi^{\alpha}(\vec{x}))}
\label{invariance}
\end{equation}

\vspace{0.1 in}

The method presented in this  section will be used to solve the problem of exclusion of double occupancy in the next chapters.

\vspace{0.3 in}

\section{THE $t-J$ MODEL FOR CORRELATED ELECTRONS} 

\vspace{0.3in}

In this section we will present the model  for the high $T_{c}$ Superconductors. We will show that the ground state can be computed  using the method of First Class constraints \cite{Dirac}.

 For the case that the hopping parameter $t$ and the one site repulsion $U$ obey the condition
$(t/U)<1$, the double occupation states  are projected out and  one obtains an  effective hamiltonian   $H \Rightarrow PH_{0}P+\frac{PH_{0}(1-P)(1-P)H_{0}P}{U}+...$ where   $P=1-n_{\uparrow}n_{\downarrow}$ is the   projection operator.  The effective model is given by:

\begin{equation}
H=-t\sum_{\vec{x},\vec{a}}\sum_{\sigma=\uparrow,\downarrow}\mathbf{\psi}^{\dagger}_{\sigma}(\vec{x})\mathbf{\psi}_{\sigma}(\vec{x}+\vec{a})+h.c.+\delta H \equiv H_{0}+\delta H
\label{free}
\end{equation}

where $\vec{x}$ represents the lattice points and $\vec{a}$ runs over to the nearest neighbor sites. This is  the $t-J$ model   where  $H_{0}$ is the hopping hamiltonian   which acts on a restricted Hilbert space where   double occupancy is excluded, while $\delta H$ represents the exchange  hamiltonian  controlled by  $J\propto \frac{t^2}{U}$.

  The Many Body ground state   is given by   $|\Psi^{0}>$.
The  exclusion of double occupancy on  each lattice point  $x$  is imposed by the    constraint condition which determines  the ground state  $|\Psi^{0}>$: 
 
\begin{equation}
\mathbf{\psi}_{\uparrow}(\vec{x})\mathbf{\psi}_{\downarrow}(\vec{x})
|\Psi^{0}>=0
\label{constrained}
\end{equation}
 
We define the constraint field  
 $Q(\vec{x})=\mathbf{\psi}_{\uparrow}(\vec{x})\mathbf{\psi}_{\downarrow}(\vec{x})$.  
To   find the ground state  $|\Psi^{0}>$   of the hamiltonian   $H_{0}+\delta H$ which obeys the constraints we have   to  satisfy the  following equations:   
 
\begin{equation}
H|\Psi^{0}>=E|\Psi^{0}> \hspace{0.1 in}  and \hspace{0.1 in}
  Q(\vec{x})|\Psi^{0}>=0
\label{eigenvalues}
\end{equation}

\vspace{0.1 in}

Once the ground state $|\Psi^{0}>$ is found,  the excitations spectrum is obtained by applying  the creation operators on this ground state such that no double occupied excited states are created.

It is easy to see that  the solution to eq. $(28)$ can be written  in the form:

\begin{equation}
|\Psi^{0}>={\Pi}_{\vec{x}}[1-Q^+(\vec{x})Q(\vec{x})]\widetilde{|\Psi^{0}>}
\label{muthu}
\end{equation}

Where  $\widetilde{|\Psi^{0}>}$ is a state which  must be determined!
The  state  $\widetilde{|\Psi^{0}>}$  belongs to the  Gutzwiller class  \cite{Gutz} and is often determined   by variational methods \cite{Muthu}. Two  possible choices for $\widetilde{|\Psi^{0}>}$ have been considered: the  $BCS$ wave function    has been used to compute the   $RVB$ state and the $Fermi$ $Liquid$ ground  state    $|F.S>=\prod_{\vec{K}=0}^{\vec{K}_{F}} \psi^{+}_{\uparrow}(\vec{K})\psi^{+}_{\downarrow}(\vec{K})|0>$  has been introduced to compute the $strongly$ $correlated$   $metallic$ $state$.

In the present paper we will  not perform a  variational calculation, instead we will  compute the ground state wave function  by determining the  additional  conditions   which the state  $|\Psi^{0}>$ has to satisfy.  We will show that using  the  Fermi surface as a the unperturbed ground state   we find  two additional constraints, the pair creation constraint  $Q^+(\vec{x})$ and the hole number constraint  $Q_{3}(\vec{x})=1-n_{\uparrow}(\vec{x})-n_{\downarrow}(\vec{x})$. The set of the three  non-commuting constraints replace the  delta function measure (which results from the condition   $Q(\vec{x})|\Psi^{0}>=0$ )  by a non-linear  integration measure . It is this measure which generates an effective temporal interaction with a time dependent coupling constant.

These results will be derived in the next chapter    using the method of first class constraints  introduced   in chapter $II$.

\vspace{0.3 in }

\section{THE APPLICATION OF FIRST CLASS CONSTRAINTS TO THE PROBLEM OF EXCLUSSION OF DOUBLE OCCUPANCY}

\vspace{0.3 in}

In this chapter  we will apply the general theory  for First Class constraints   introduced  in chapter    II to solve the model presented in chapter III.  
The eigenfunction $|\Psi^0>$ must obey $ Q(\vec{x})|\Psi^0>=0$.
We will show that away from half fillings we can use a set of first class constraints  defined only for negative energies $Q^{(-)}_{\alpha}(\vec{x})$ where $\alpha=1,2,3$. 
The  commutator $[ Q^{(-)}(\vec{x}),H]$ generates   secondary constraints   $q^{(-)}_{r}(\vec{x})$, $r=1,2,3$  which will be neglected  for describing the low energy physics. The justification for this approximation is given in the appendix of this paper where we show that the effective action generated by the secondary constraints are irrelevant  according to the R.G. analysis at low energies.

Due to the constraints, the non interacting Fermi energy will be shifted by $\delta{\mu_{F}}$ to a new  value. The   metallic  behavior will be characterized  by the vanishing of the  renormalized  chemical potential shift $[\delta{\mu_{F}}]_{R}=0$ \cite{Shankar}.

At half fillings we have two additional constraints: $ Q^{\dagger}(\vec{x})$ and the hole number operator 
 $Q_{3}(\vec{x}) \equiv 1-[\mathbf{\psi}^{\dagger}_{\uparrow}(\vec{x})\mathbf{\psi}_{\uparrow}(\vec{x})+\mathbf{\psi}^{\dagger}_{\downarrow}(\vec{x}) \mathbf{\psi}_{\downarrow}(\vec{x})]$.
  At  half fillings the three constraints satisfy:  $ Q(\vec{x})|\Psi^0>=0$,  $Q^{\dagger}(\vec{x})|\Psi^0>=0$ and $Q_{3}(\vec{x})|\Psi^{0}>=0$.  
$Away$ $from$ $half$ $fillings$ we have only one first class constraints  $Q(\vec{x})$,  the other two constraints are neither first class nor second class. This difficulty can be resolved by modifying the constraints.
Away  from half fillings we will restrict the constraints  only to  negative energies /holes excitations    $Q^{(-)}_{\alpha}(\vec{x})$ where $\alpha=1,2,3$.
We introduce the definitions:

\begin{equation}
\mathbf{\psi}_{\sigma}(\vec{x})=\mathbf{\psi}^{(+)}_{\sigma}(\vec{x})+\mathbf{\psi}^{(-)}_{\sigma}(\vec{x}) ;\mathbf{\psi}^{\dagger}_{\sigma}(\vec{x})=\mathbf{\psi}^{\dagger (+)}_{\sigma}(\vec{x})+\mathbf{\psi}^{\dagger (-)}_{\sigma}(\vec{x})
\label{psi}
\end{equation}

The notation  $(+)$ represents the particles excitations for  $positive$ energies     and  $(-)$ describes the holes excitations  for $negative$  energies, both measured with respect to the   renormalized Fermi energy.

$ \mathbf{\psi}^{(+)}_{\sigma}(\vec{x})$  and 
$ \mathbf{\psi}^{\dagger(-)}_{\sigma}(\vec{x})$ act as a  destruction operators with respect to the ground state $|\Psi^{0}>$ and obey   
$ \mathbf{\psi}^{(+)}_{\sigma}(\vec{x})|\Psi^{0}>= \mathbf{\psi}^{\dagger (-)}_{\sigma}(\vec{x})|\Psi^{0}>=0$. From  \cite{Fetter}  we learn that   the field $ \mathbf{\psi}^{(-)}_{\sigma}(\vec{x})$ is build from  the Fourier momentum components   $\vec{K}\leq \vec{K}_{F}$, where   $\vec{K}_{F}$ corresponds to the non-interacting Fermi momentum. Similarly $\mathbf{\psi}^{(+)}_{\sigma}(\vec{x})$ is build from  the Fourier  momentum components  $\vec{K}> \vec{K}_{F}$.
The effective renormalization of the Fermi surface caused by the constraints is included into the chemical potential shift $\delta{\mu_{F}}$, which will be computed within the R.G. calculations.

 The  renormalized ground state $|\Psi^{0}>$  has  no particles above $\vec{K}_{F}$  and no holes below  $\vec{K}_{F}$. (At this step we do not make any assumption  about the nature of the  discontinuity  of the Fermi Surface  (the occupation number at $\vec{K}_{F}$).
 
 Acting with the constraint operator $Q(\vec{x})$ on the ground state $|\Psi^{0}>$ we observe:
 
\begin{equation} Q(\vec{x})|\Psi^{0}>=(\mathbf{\psi}^{(+)}_{\uparrow}(\vec{x})+\mathbf{\psi}^{(-)}_{\uparrow}(\vec{x}))(\mathbf{\psi}^{(+)}_{\downarrow}(\vec{x})+\mathbf{\psi}^{(-)}_{\downarrow}(\vec{x}))|\Psi^{0}> =\mathbf{\psi}^{(-)}_{\uparrow}(\vec{x}))\mathbf{\psi}^{(-)}_{\downarrow}(\vec{x})|\Psi^{0}>=0 
\label{minus}
\end{equation}

This equation shows that the constraints can  be restricted to the holes type excitations $
Q^{(-)}(\vec{x})=\mathbf{\psi}^{(-)}_{\uparrow}(\vec{x}))\mathbf{\psi}^{(-)}_{\downarrow}(\vec{x})$. For  positive energies  the constraint is automatically satisfied,  therefore   the constraint $Q(\vec{x})$ is restricted to 
$Q^{(-)}(\vec{x})$.

Using the holes  representation, $\mathbf{\psi}^{(-)}_{\sigma}(\vec{x})$ and $\mathbf{\psi}^{\dagger (-)}_{\sigma}(\vec{x})$ we construct the  new representations  for the constraints:

\begin{eqnarray} Q^{(-)}(\vec{x})&=&\mathbf{\psi}^{(-)}_{\uparrow}(\vec{x})\mathbf{\psi}^{(-)}_{\downarrow}(\vec{x})\label{eq:s1}\\
Q^{\dagger (-)}(\vec{x})&=&[\mathbf{\psi}^{(-)}_{\uparrow}(\vec{x})\mathbf{\psi}^{(-)}_{\downarrow}(\vec{x})]^{\dagger}\label{eq:s2}\\
Q_{3}^{(-)}(\vec{x})&=&1-[\mathbf{\psi}^{\dagger (-)}_{\uparrow}(\vec{x})\mathbf{\psi}^{(-)}_{\uparrow}(\vec{x})+\mathbf{\psi}^{\dagger(-)}_{\downarrow}(\vec{x}) \mathbf{\psi}^{(-)}_{\downarrow}(\vec{x})] \label{eq:s3}
\end{eqnarray}  

It is convenient  to  work with real constraints. We introduce    $two$ $real$ $constraints$, $Q^{(-)}_{1}(\vec{x})$ and $Q^{(-)}_{2}(\vec{x})$ using  the $constraint$ $Q^{(-)}(\vec{x})$ and  $Q^{\dagger(-)}(\vec{x})$. 
 
 \begin{eqnarray}
Q^{(-)}_{1}(\vec{x})&=&\frac{1}{\sqrt{2}}(Q^{(-)}(\vec{x})+Q^{\dagger (-)}(\vec{x})\label{eq:g1}\\
  Q^{(-)}_{2}(\vec{x})&=&\frac{1}{\sqrt{2}i}(Q^{(-)}(\vec{x})-Q^{\dagger (-)}(\vec{x})) 
\end{eqnarray}

In addition to this two real  constraints we have the third real constraint  $Q^{(-)}_{3}(\vec{x}) $  given by equation $(34)$.
The set of the hole type   constraints  $Q^{(-)}_{\alpha}(\vec{x})|\Psi^{0}>$ ,$\alpha=1,2,3$ represent the   Non-Abelian First Class constraints for  our problem. The commutators of  the constraints  obey:

\begin{equation}
[Q^{(-)}_{\alpha}(\vec{x}),Q^{(-)}_{\beta}(\vec{x})]=if_{\alpha,\beta}^{\gamma}Q^{(-)}_{\gamma}(\vec{x})
\label{su(2)}
\end{equation}
The $Non$-$Abelian$ $First$ $Class$ $constraints$  are characterized  by  the $structure$ constants   $f_{\alpha,\beta}^{\gamma}$, 
 $f_{\alpha,\beta}^{\gamma}=-f_{\beta,\alpha}^{\gamma}=1$  for $ \alpha \neq \beta\neq \gamma $ and zero otherwise.
 
  The commutator of the constraints with the hamiltonian $H$ satisfies the equation:
  
$[Q^{(-)}_{\alpha}(\vec{x}),H]\approx T^{\alpha}_{\beta} (\vec{x})Q^{(-)}_{\beta}(\vec{x})$, 
where $T^{\alpha}_{\beta}(\vec{x})$ stands for a set of local matrix operators. The symbol $\approx$ means that the derivatives of the constraints operators  have been neglected. This approximation  can be seen by computing  the commutator of the constraint with the kinetic energy:

  $[\mathbf{\psi}_{\sigma=\downarrow}(\vec{x})\mathbf{\psi}_{\sigma=\uparrow}(\vec{x}),H_{0}]=t\sum_{\vec{a}}[\mathbf{\psi}_{\sigma=\downarrow}(\vec{x})\mathbf{\psi}_{\sigma=\uparrow}(\vec{x}+\vec{a})+\mathbf{\psi}_{\sigma=\downarrow}((\vec{x}+\vec{a}))\mathbf{\psi}_{\sigma=\uparrow}(\vec{x})]$
 
The new secondary  constraints are obtained by taking the difference of the commutator  $\frac{1}{t}[Q(\vec{x}),H_{0}]$   with the  primary first class  constraints    $Q(\vec{x})$: 

\vspace{0.1 in}

$q(\vec{x})\equiv \sum_{\vec{a}}[\mathbf{\psi}_{\sigma=\downarrow}(\vec{x})\mathbf{\psi}_{\sigma=\uparrow}(\vec{x}+\vec{a})+\mathbf{\psi}_{\sigma=\downarrow}((\vec{x}+\vec{a}))\mathbf{\psi}_{\sigma=\uparrow}(\vec{x})] -Q(\vec{x})$

\vspace{0.1 in}

$q^{+}(\vec{x})\equiv \sum_{\vec{a}}[\mathbf{\psi}_{\sigma=\downarrow}(\vec{x})\mathbf{\psi}_{\sigma=\uparrow}(\vec{x}+\vec{a})+\mathbf{\psi}_{\sigma=\downarrow}((\vec{x}+\vec{a}))\mathbf{\psi}_{\sigma=\uparrow}(\vec{x})]^{+} -Q^{+}(\vec{x})$

and 
\vspace{0.1 in}

$q_{3}(\vec{x})\equiv \sum_{\vec{a}}[\mathbf{\psi}_{\sigma=\downarrow}(\vec{x})\mathbf{\psi}_{\sigma=\downarrow}(\vec{x}+\vec{a})+\mathbf{\psi}_{\sigma=\uparrow}((\vec{x}+\vec{a}))\mathbf{\psi}_{\sigma=\uparrow}(\vec{x})]  -[Q_{3}(\vec{x})-1]$

Using the linear transformation given by equations $(35)$ and $(35)$ we obtain the $new$ set of $secondary$ $first$  $class$ $real$ $constraints$:  $q^{(-)}_{r}(\vec{x})$, $r=1,2,3$.

 The   secondary first class  constraints  modify   the  Lagrangian $L$ by $\delta L$ :

    $\delta L =\sum_{r=1}^{3}\hat{\pi}_{r}(\vec{x},t)\partial_{t}\hat{\lambda}^{r}(\vec{x},t)-\sum_{r=1}^{3}(\hat{\pi}_{r}(\vec{x},t)\hat{\chi}^{r}(\vec{x},t)+\hat{\lambda}^{r}(\vec{x},t)q^{(-)}_{r}(\vec{x},t))  $
    
where  $\hat{\lambda}^{r}(\vec{x},t)$   are the Lagrange multipliers which enforce the secondary  constraints and $\hat{\pi}_{r}(\vec{x},t)$ are  the canonical momentum conjugated to the new Lagrange  multipliers $\hat{\lambda}^{r}(\vec{x},t)$. 
The integration over the new Lagrange multipliers will generate   products of  $q^{(-)}_{r}(\vec{x})$ pairs.  Such operators will induce  effective interactions  which have  the dimensions  of  the square of the kinetic energy  with  two  time integrations.  According to the analysis  given in the Appendix  the engineering  dimensions    for such operators is $-1$ and therefore they  are strongly irrelevant at low energies ( in comparison to the engineering dimensions  $1$ for  the effective interaction  due to the  primary constraints). Therefore  we will ignore the  Lagrangian  induced by   the secondary  constraints  or/and  higher order  constraints generated by the commutators   $[q^{(-)}_{r}(\vec{x}),H_{0}]$, $r=1,2,3$  (such operators     contain higher  order derivatives in comparison with the   constraints $q^{(-)}_{r}(\vec{x})$).

The canonical phase space action  for the  holes constraints which  replaces  eq.$(8)$ (without the generated constraints  described by $\delta L$)  is  given by:
 
\begin{equation}
L=\sum_{\sigma=\uparrow,\downarrow}i\hbar\psi^{\dagger}_{\sigma}(\vec{x},t)\partial_{t}\psi_{\sigma}(\vec{x},t)+\sum_{\alpha=1}^{3}\pi_{\alpha}(\vec{x},t)\partial_{t}\lambda^{\alpha}(\vec{x},t)-h(\vec{x},t)-\sum_{\alpha=1}^{3}(\pi_{\alpha}(\vec{x},t)\chi^{\alpha}(\vec{x},t)+\lambda^{\alpha}(\vec{x},t)Q^{(-)}_{\alpha}(\vec{x},t))
\label{action}
\end{equation}

where $\pi_{\alpha}(\vec{x})$ is  the canonical  momentum conjugate to the Lagrange multiplier  $\lambda^{\alpha}(\vec{x})$.  This action has two sets of constraints $Q^{(-)}_{\alpha}(\vec{x})|\Psi^{0}>=0$ and  
$\pi_{\alpha}(\vec{x})|\Psi^{0}>=0$.  The Lagrange multipliers are:  $\lambda^{\alpha}(\vec{x})$ and the unknown one $\chi^{\alpha}(\vec{x})$.
Using the general theory for First Class constraints presented in section II, we  will express the action in eq. $(38)$ using new anti-commuting   fields. The   anti-commuting fields $C^{\alpha}(\vec{x})$ and $b_{\alpha}(\vec{x})$  are used to replace  the  holes constraints  $Q^{(-)}_{\alpha}(\vec{x},t) $ and  the anti-commuting fields  $e^{\alpha}(\vec{x})$ and $f_{\alpha}(\vec{x})$ replace the  momenta  constraints $\pi_{\alpha}(\vec{x})$. The   unknown Lagrange multiplies    $\chi^{\alpha}$  act as gauge fixing conditions.

  We replace the state $|\Psi^{0}>$ by  the state $|\Psi>$  defined for  the extended Hilbert with the extended constraint operator  $\Omega $ for the extended Hilbert space (see eq.$(19)$),
 $\Omega =\Omega _{0}+\Omega_{NL}$   where  $\Omega_{NL}$ is chosen such that the condition  $\Omega^2=0$ is satisfied.
 
  The operator 
$\Omega _{0}$  (see eq.$(20)$) is given in terms of the $physical$ $constraints$  $Q^{(-)}_{\alpha}(\vec{x})$ and $\pi_{\alpha}(\vec{x})$:

\begin{equation}
\Omega_{0}
=\int d^dx\left[\sum_{\alpha=1}^{3}C^{\alpha}(\vec{x})Q^{(-)}_{\alpha}(\vec{x})+\sum_{\alpha=1}^3f^{\alpha}(\vec{x})\pi_{\alpha}(\vec{x})\right]
\label{ome}
\end{equation}

The Nonlinear operator  $\Omega_{NL}$  (see eq.$(21)$)is obtained using the structure constants given in eq.$(37)$.

\begin{equation}
\Omega_{NL}(\vec{x})=\sum_{\alpha=1}^3 C^{\alpha}(\vec{x})\left[\int \,d^{d}x^{1}\sum_{\beta=1}^3 \sum_{\gamma=1}^3 \frac{-i}{2}C^{\gamma}(\vec{x}) f_{\alpha,\beta}^{\gamma}b_{\beta}(\vec{x}^{1})]\right]
\label{NLA}
\end{equation}

According to the theorem \cite{Vilkov}  the path integral 
 is invariant under the change of  the Lagrange multipliers  $\chi^{\alpha}(\vec{x})$.  For the continuation of this article  we will  choose   $\chi^{\alpha}(\vec{x})=0$ and obtain from eq.$(23)$ the Fermionic operator $W$.

\begin{equation}
W=\int d^{d} x\sum_{\alpha=1}^{3}b_{\alpha}(\vec{x})\lambda^{\alpha}(\vec{x})
\label{ccc}
\end{equation}

We compute the anti-commutator  $\frac{i}{\hbar}[W,\Omega]_{+}$ and find:
\begin{equation}
\frac{i}{\hbar}[W,\Omega]_{+}=\sum_{\alpha=1}^{3}\lambda^{\alpha}(\vec{x})Q^{(-)}_{\alpha}(\vec{x})+i\hbar \sum_{\alpha=1}^{3}b_{\gamma}(\vec{x})f^{\alpha}(\vec{x}) +\sum_{\alpha=1}^{3}\sum_{\beta=1}^{3}\sum_{\gamma=1}^{3}\frac{1}{\hbar}C^{\beta}(\vec{x})\lambda^{\alpha}(\vec{x},t)f_{\alpha,\beta}^{\gamma}b_{\gamma}(\vec{x},t)
\label{ffc}
\end{equation}

This result allows to obtain the effective hamiltonian in the enlarged  space:

$H_{effective}=H_{c}+[\Omega,W]_{+}$   where $H_{c}$  is chosen such that $[H_{c},\Omega]=0$.

In this  case the   difference between  $H_{c}$  and $H$ is given by operators which contain higher order derivatives and therefore are irrelevant for  low energy Physics.

 Using the effective hamiltonian $H_{effective}=H+[\Omega,W]_{+}$  we obtain the action $S$ as  a function of the initial and final times  $ t_{i}$ and  $ t_{f}$,  $T= t_{i}-t_{f}$ .

\begin{eqnarray}
S[t_{i},t_{f}]&=&\int_{t_{i}}^{t_{f}} \,dt[\sum_{\sigma=\downarrow,\uparrow}(-i\hbar)\psi _{\sigma}^{\dagger }(x,t)\partial_{t}\psi_ {\sigma}(x,t)+\sum_{\alpha=1}^{3}\pi_{\alpha }( x,t) \partial
_{t}\lambda^{\alpha }( \vec{x},t)+\nonumber \\ && \sum_{\alpha=1}^{3}\partial _{t}C^{\alpha}(\vec{x},t)b_{\alpha}(x,t)+\sum_{\alpha=1}^{3}\partial _{t}e_{\alpha}(\vec{x},t)f^{\alpha}(x,t)-  \nonumber \\ && [h(\vec{x},t)+ \sum_{\alpha=1}^{3} \lambda^{\alpha}(\vec{x},t)Q^{(-)}_{\alpha}(\vec{x},t)+ i\hbar\sum_{\alpha=1}^{3} b_{\alpha}(\vec{x},t)f^{\alpha}(\vec{x},t)+ \nonumber \\&& i \sum_{\alpha=1}^{3}\sum_{\beta=1}^{3}\sum_{\gamma=1}^{3}\frac{1}{\hbar}C^{\beta}(\vec{x})\lambda^{\alpha}(\vec{x},t)f_{\alpha,\beta}^{\gamma}b_{\gamma}(\vec{x},t)]]
\end{eqnarray}

For commuting constraints, the last term in eq.$(43)$ vanishes (the anti-commuting fields are decoupled). For such a situation the constraints in eq.$(43)$ can be treated by delta function constraints.
 Once the non-commutativity of the constraints is considered one generate a non-linear measures for the Lagrange multipliers (due to the last term in eq.$(43)$. 
  
 The full Non-Abelian action  given in    equation $(43)$ describes a  canonical evolution in an enlarged Hilbert space.  The  physical observables  are   obtained only after  we  project  the  extended state into the physical Hilbert space.
This is achieved  by imposing  temporal boundary conditions on the non physical anti-commuting  coordinates $C^{\alpha}(x,t)$ ,$e_{\alpha}(x,t)$ and the conjugate  momenta $\pi_{\alpha}(\vec{x},t)$ ,$\alpha=1,2,3$. 
\begin{eqnarray}
C^\alpha(x,t_{i})=C^\alpha(x,t_{f})=
e_{\alpha}(x,t_{i})=e_{\alpha}(x,t_{f})= 
\pi_{\alpha}(\vec{x},t_{i})=\pi_{\alpha}(\vec{x},t_{f})=0
\end{eqnarray}

  The projection into the physical space is achieved  by the use of initial time $t_{i}$ and final time $t_{f}$ boundary conditions  \cite{Teitelboim}. 
 
\begin{equation}
U_{phys}[\overline{\psi}_{f}(t_{f});\psi_{i}(t_{i})]\equiv U[\overline{\psi}_{f}(t_{f}),\pi_{\alpha}(x,t_{f})=0,C^\alpha(x,t_{f})=0,e_{\alpha}(x,t_{i})=0;\psi_{i}(t_{i})]
\label{bc}
\end{equation}

The physical evolution matrix elements  are  given by:

\begin{equation}
U_{phys}[\overline{\psi}_{f}(t_{f});\psi_{i}(t_{i})]= \int D(\overline{\psi}_{\sigma},\psi_{\sigma})D(C^{\alpha},b_{\alpha})D(e_{\alpha},f^{\alpha})
D(\lambda^{\alpha},\pi_{\alpha}) exp[\frac{1}{2}(\overline{\psi}_{f}\psi_{f}+\overline{\psi}_{i}\psi_{i}+\frac{i}{\hbar}S[t_{i},t_{f}]]
\label{conu}
\end{equation}

Where $S[t_{i},t_{f}]$ is the action in equation $(43)$ . $D(\overline{\psi}_{\sigma},\psi_{\sigma})$ represents the  Grassmann measure,
$D(C^{\alpha},b_{\alpha})$ and $D(e_{\alpha},f^{\alpha})$ represents the integration measure \cite{Teitelboim} for the fictitious non-commuting  degrees of freedom.  $ D(\lambda^{\alpha},\pi_{\alpha})$ is  the phase space integration over the Lagrange multipliers and the canonical conjugated variables.
 The action  $S[t_{i},t_{f}]$ can be written as a sum of the $physical$ $action$ $\widetilde{S}[t_{i},t_{f}]$ and a  $non-physical$ $action$   $S^{ghost}[t_{i},t_{f}]$ given in terms of the  non-commuting  degrees  of freedom $C^{\alpha}$,$b_{\alpha }$ and $e_{\alpha}$ , $f^{\alpha}$:
 
\begin{equation}
S[t_{i},t_{f}]\equiv \widetilde{S}[t_{i},t_{f}]+S^{ghost}[t_{i},t_{f}]
\label{ssum}
\end{equation}

For fixed values of $t_{i}$ and $t_{f}$  we integrate over the  $non-physical$ non-commuting  degrees  of freedom $C^{\alpha}$,$b_{\alpha }$ and $e_{\alpha}$ , $f^{\alpha}$. We obtain an effective action $S^{eff.}[\lambda^{\alpha};t_{i},t_{f}]$ as a function of the time interval and the Lagrange multipliers.  This action is obtained once the non-physical fermions fields have been integrated out and the boundary conditions for the conjugate momenta $\pi_{\alpha}(t_{i})=\pi_{\alpha}(t_{f})=0$  have been used.

\begin{equation}
e^{\frac{i}{\hbar}S^{eff.}[\lambda^{\alpha};t_{i},t_{f}]}\equiv \int \,D(C^{\alpha},b_{\alpha})D(e_{\alpha},f^{\alpha})e^{\frac{i}{\hbar}S^{ghost}[t_{i},t_{f}]}|_{\pi_{\alpha}(t_{i})=\pi_{\alpha}(t_{f})=0}
\label{ghost}
\end{equation}

This allows to represent the matrix elements of the physical evolution operator by the formula:
\begin{eqnarray}
U_{phys}[\overline{\psi}_{f}(t_{f});\psi_{i}(t_{i})]&=& \int
D(\overline{\psi}_{\sigma},{\psi}_{\sigma}) D(\lambda^{\alpha})e^{\frac{i}{\hbar}S^{eff.}[\lambda^{\alpha};t_{i},t_{f}]}e^{[\frac{1}{2}(\overline{\psi}_{f}\psi_{f}+\overline{\psi}_{i}\psi_{i})+\frac{i}{\hbar}\widetilde{S}[t_{i},t_{f}]]} =
\nonumber \\&&\int
D(\overline{\psi}_{\sigma},{\psi}_{\sigma}) \mathcal{D}(\lambda^{\alpha};t_{i}-t_{f})e^{[\frac{1}{2}(\overline{\psi}_{f}\psi_{f}+\overline{\psi}_{i}\psi_{i})+\frac{i}{\hbar}\widetilde{S}[t_{i},t_{f}]]} 
\end{eqnarray}

where $\mathcal{D}\lambda(\vec{x};t_{i}-t_{f})$ is a $time$ $dependent$  $non-linear$ $measure $ obtained by integrating out the non-physical anti-commuting  fields which    describe the dynamics  of the canonical conjugate fields. The  measure is  defined trough the path integral in eqs. $(48)$-$(49)$. As a result any physical Green's function will depend on a coupling constant which is a function of the time interval $T= t_{i}-t_{f}$  (generated by the projection of the non physical Fermionic degrees of freedom at the initial and final time).

\vspace{0.3 in}

\section{COMPUTATION OF THE INTEGRATION MEASURE   $\mathcal{D}(\lambda^{\alpha};t_{i}-t_{f})$}

\vspace{0.3 in}

The physical evolution operator given in equation  $(49)$  depends on the integration measure $\mathcal{D}(\lambda^{\alpha};t_{i}-t_{f})$. This measure is  computed from  the  action $S^{ghost}[t_{i},t_{f}]$ and is defined by   the equations $(43)$, $(47)$-$49$. The explicit form of $S^{ghost}[t_{i},t_{f}]$ is given by:
\begin{eqnarray}
\frac{i}{\hbar}S^{ghost}[t_{i},t_{f}]&=&\int_{t_{i}}^{t_{f}}[ \sum_{\alpha=1}^{3}\partial _{t}C^{\alpha}(x,t)b_{\alpha}(x,t)+\sum_{\alpha=1}^{3}\partial _{t}e_{\alpha}(x,t)f^{\alpha}(x,t)-\sum_{\alpha=1}^{3} b_{\alpha}(x,t)f^{\alpha}(x,t)+ \nonumber \\&& \sum_{\alpha=1}^{3}\sum_{\beta=1}^{3}\sum_{\gamma=1}^{3}\frac{1}{\hbar}C^{\beta}(\vec{x})\lambda^{\alpha}(x,t)f_{\alpha,\beta}^{\gamma}b_{\gamma}(x,t)]\,dt
\end{eqnarray}

The  integration  over the non-physical fermionic field is done  using  the variation of the  action  $\delta[S^{ghost}[[t_{i},t_{f}]]=0$. We obtain the relation
 $\partial_{t}(e_{\alpha}(x,t))=b_{\alpha}(x,t)$. 
 As a result we find that the  integration measure is given by,  
 $\mathcal{D}(\lambda^{\alpha};t_{i},t_{f})= D(\lambda^{\alpha})e^{\frac{i}{\hbar}S^{eff.}[\lambda^{\alpha};t_{i},t_{f}]}$. 
The effective action  
$e^{\frac{i}{\hbar}S^{eff.}[\lambda^{\alpha};t_{i},t_{f}]}$  is determined by the Grasmann  integration:

\begin{eqnarray}
e^{\frac{i}{\hbar}S^{eff.}[\lambda^{\alpha};t_{i},t_{f}]}&=&\int D(C^{\alpha})D(e_{\alpha})e^{[\int_{t_{i}}^{t_{f}}\,dt[C^{\alpha}(\vec{x},t)(-I_{(\alpha,\beta)}\frac{d^2}{dt^2}-\frac{1}{\hbar}f_{\alpha,\beta}^{\gamma}\lambda^\beta (\vec{x},t)\frac{d}{dt})e_{\beta}(\vec{x},t)]}=\nonumber \\ &&
Det[-\frac{d^2}{dt^2}]Det[-I\frac{d^2}{dt^2}-\frac{1}{\hbar}f_{\alpha,\beta}^{\gamma}\lambda^\beta (\vec{x},t)\frac{d}{dt})]
\end{eqnarray}

with $I=I_{(\alpha,\beta)}$ being the identity matrix.
Following \cite{Naka,Brian} we evaluate the two  determinants in eq. $(51)$.
The determinants are a function of the time intervals $T=t_{i}-t_{f}$. For convergence  reasons at $T\rightarrow\infty$    we  $rotated$ the  $time$ $contour$ \cite{Greiner}. We define a new time interval 
$\widehat{T}=T e^{i\delta}$.  
The effective action $S^{eff.}[\lambda^{\alpha};t_{i}e^{i\delta},t_{f}e^{i\delta}]\equiv S^{eff.}[\varphi(\vec{x},\widehat{T});t_{i}e^{i\delta},t_{f}e^{i\delta}]$ is given by,
\begin{equation}
S^{eff.}[\varphi(\vec{x},\widehat{T});\widehat{t_{i}},\widehat{t_{f}}]=\Lambda^d\int\,d^{d}x Log[\pi \frac{Sin(\varphi(\vec{x},\widehat{T}))}{\varphi(\vec{x},\widehat{T})}]
\label{rotated}
\end{equation}

where $\Lambda$ is the ultraviolet cut-off and  $d$ is the space dimension.  The integration variable $\lambda^{\alpha}(\vec{x})$ is  replaced by $\varphi^{\alpha}(\vec{x},\widehat{T})$ given by  
$\varphi^{\alpha}(\vec{x},\widehat{T})= \widehat{T}\frac{\lambda^{(\alpha)}(\vec{x})}{\hbar}$ with the amplitude
$\varphi(\vec{x},\widehat{T})\equiv \widehat{T} \sqrt{(\frac{\lambda^{(1)}(\vec{x})}{\hbar})^2+(\frac{\lambda^{(2)}(\vec{x})}{\hbar})^2+(\frac{\lambda^{(3)}(\vec{x})}{\hbar})^2}$. Eq.$(52)$ represents the new integration  $measure$ for  the Lagrange multipliers.

\vspace{0.3 in}

\section{THE  EFFECTIVE  MODEL  FOR  EXCLUSION OF DOUBLE OCCUPANCY IN TWO DIMENSIONS} 

\vspace{0.3 in}

The action in equation $(43)$ can be written in a simplified form once we replace the measure 
 $d\lambda^{\alpha}(\vec{x},t)$  with the non-linear measure $d \lambda^{\alpha}(\vec{x})e^{Log[\pi \frac{Sin(\varphi(\vec{x},\widehat{T}))}{\varphi(\vec{x},\widehat{T})}]}$.
 
\vspace{0.1 in}

$S[t_{i},t_{f}]=\int_{t_{i}}^{t_{f}} \,dt[\sum_{\sigma=\downarrow,\uparrow}(-i\hbar)\psi _{\sigma}^{\dagger }(x,t)\partial_{t}\psi_ {\sigma}(x,t)-
[h(\vec{x},t)+ \sum_{\alpha=1}^{3} \lambda^{\alpha}(\vec{x},t)Q^{(-)}_{\alpha}(\vec{x},t)]]$

 The new  measure   $d \lambda^{\alpha}(\vec{x})e^{Log[\pi \frac{Sin(\varphi(\vec{x},\widehat{T}))}{\varphi(\vec{x},\widehat{T})}]}$ was obtained as a result of the non-commuting constraints  $Q^{(-)}_{\alpha}(\vec{x},t)$.
The  change in the measure   can be rewritten explicitly in terms of the effective interaction  given in equation $(52)$.

\vspace{0.3 in}

\section{THE EFFECTIVE MODEL IN THE ABSENCE OF THE EXCHANGE INTERACTION} 

\vspace{0.3 in}

In this section we will present the effective model for a free electron system which obeys  the exclusion of double occupancy. 

 We will consider the case where  the  exchange  term $\delta H=0$. Therefore we will replace   $h(\vec{x},t)\rightarrow h_{0}(\vec{x},t)$   (in  equation $(43)$).
We believe that this model describes the situation at $zero$ $temperatures$ away from half fillings  above certain holes concentration $x_{c}$ (where  the magnetic order is absent).
When  the exchange is  zero we will not be able to investigate the metal insulator transition, which takes place  close to the half filled case $x\rightarrow 0$. We will  consider the case where the  holes density $x$ obeys $x>x_{c}$. Experimentally we know that at T=0 such a window exists between the magnetic ordered state and the appearance of superconductivity.
In order to study  the nature of the  metallic state  at zero temperature and holes densities  $x>x_{c}$ it is reasonable  to suppress the superconducting order  by restricting the exchange interaction to  zero.  We will show that at low holes concentrations $x_{c}<x<<1$ the imaginary part of the self energy is proportional to $\omega$. For large holes densities $x\rightarrow 1$,  a crossover transition to a Fermi liquid  with   the imaginary part of the self energy  proportional to $\omega^2$ is obtained.

 In order to perform a R.G.  study we use the  $free$ Fermi liquid representation  in two dimensions. 

$\mathcal{S^{0}}[\widehat{t_{i}},\widehat{t_{f}}]=\int_{0}^{\pi} \frac{ds}{\pi}\int_{\widehat{t_{i}}}^{\widehat{t_{f}}}\,dt \int\frac{d\epsilon}{2\pi}J[\epsilon,s]\sum_{\sigma=\uparrow,\downarrow}[R^{\dagger}_{\sigma}(t,\epsilon;s)(i\hbar \partial _{t}-\hbar\epsilon)R_{\sigma}(t,\epsilon;s) +L^{\dagger}_{\sigma}(t,\epsilon;s)(i\hbar \partial _{t}+\hbar\epsilon)L_{\sigma}(t,\epsilon;s)$

where $R_{\sigma}(t,\epsilon;s)$  and $L_{\sigma}(t,\epsilon;s)$ are the right and left chiral fermions in the channel $s$. The polar angle  $s$ on the Fermi surface is  restricted to the region $[0-\pi]$. The momentum excitation normal to the Fermi Surface is given by $\frac{\epsilon}{v_{F}}$ where   $v_{F}$ is the Fermi velocity. $J[\epsilon,s]=\frac{|\vec{K}_{F}(s)|}{|\vec{v}_{F}(s)|}$  is the Jacobian transformation from the $K_{x}$ and $K_{y}$ coordinates to the energy $\epsilon$ and  polar angle $s$.
We will rescale the Fermionic field by $\sqrt{\int_{0}^{\pi}{\frac{ds}{\pi}J[\epsilon,s]} }$. As a result, the Fermi surface action will depend only on the $dimensionless$ Jacobian   $\widehat{J}[\epsilon,s]=\frac{J[\epsilon,s]}{\int_{0}^{\pi}\frac{ds}{\pi}J[\epsilon,s]}$.

The effect of the constraints are  supposed to shift the position of free Fermi surface  given  by the Fermi vector $\vec{K_{F}}(s)$.  This effect will be taken in consideration  by a finite shift of the chemical potential $\delta{\mu_{F}}(s)$. As a result,   the non interacting action $\mathcal{S^{0}}[\widehat{t_{i}},\widehat{t_{f}}]$ will be replaced by 
$S^{0}[\widehat{t_{i}},\widehat{t_{f}}]$. 
The complete action for our  model as obtained from the previous chapter, including the shifted Fermi Surface, is:

\begin{equation} S[\widehat{t_{i}},\widehat{t_{f}}]=S^{0}[\widehat{t_{i}},\widehat{t_{f}}]+S^{I-hole}[\widehat{t_{i}},\widehat{t_{f}}]+S^{eff.}[\varphi^{\alpha}(\vec{x},\widehat{T});\widehat{t_{i}},\widehat{t_{f}}] 
\label{efffectivep}
\end{equation}

where  $S^{eff.}[\lambda^{\alpha};\widehat{t_{i}},\widehat{t_{f}}]$  is given in equation $(52)$   and represents the modification  of  the  integration measure for  the Lagrange multipliers $\lambda^{\alpha}(\vec{x})$.
$S^{I-hole}[\widehat{t_{i}},\widehat{t_{f}}]$ which   originally was given by $\sum_{\alpha=1}^{3} \lambda^{\alpha}(\vec{x},t)Q^{(-)}_{\alpha}(\vec{x},t)$ is written in terms of the  new variables  $\varphi_{\alpha}(\vec{x},\widehat{T})$, $\alpha=1,2,3$:

\begin{equation}
S^{I-hole}[\widehat{t_{i}},\widehat{t_{f}}]=\int_{\widehat{t_{i}}}^{\widehat{t_{f}}}\,dt\int \,d^{2}x[ \sum_{\alpha=1}^{3}\frac{\varphi_{\alpha}(\vec{x},\widehat{T})}{\widehat{T}}Q^{(-)}_{\alpha}(\vec{x},t)]
\label{hok}
\end{equation}

\begin{eqnarray}
S^{0}[\widehat{t_{i}},\widehat{t_{f}}]&=&\int_{0}^{\pi} \frac{ds}{\pi}\int_{\widehat{t_{i}}}^{\widehat{t_{f}}}\,dt \int\frac{d\epsilon}{2\pi}J[\epsilon,s]\sum_{\sigma=\uparrow,\downarrow}[R^{\dagger}_{\sigma}(t,\epsilon;s)(i\hbar \partial _{t}-\hbar\epsilon)R_{\sigma}(t,\epsilon;s)\nonumber \\&& +L^{\dagger}_{\sigma}(t,\epsilon;s)(i\hbar \partial _{t}+\hbar\epsilon)L_{\sigma}(t,\epsilon;s) + \nonumber \\&&R^{\dagger}_{\sigma}(t,\epsilon;s)\delta{\mu_{F}}(s)R_{\sigma}(t,\epsilon;s)+
L^{\dagger}_{\sigma}(t,\epsilon;s)\delta{\mu_{F}}(s)L_{\sigma}(t,\epsilon;s)]
\end{eqnarray}

where $S^{0}[\widehat{t_{i}},\widehat{t_{f}}]$  represents   the  free fermion action and  $\delta{\mu_{F}}(s)$ is the  shift in the  chemical 
potential  induced by constraints. The metallic phase will be identified by the vanishing of the $renormalized$ $chemical$ 
$potential$ $shift$ $[\delta{\mu_{F}}(s)]_{R}=0$ \cite{Shankar}.

The  action in eq.$(53)$   represents the complete solution for the problem of exclusion of double occupancy for the  tight binding fermion model. It is important to mention   that the action describes 
 an effective interaction for the holes  which depends explicitly on the time interval  
$T=t_{i}-t_{f}$.  We have a situation where the Lagrange multipliers  $\lambda^{\alpha}(x)$ can be treated as  random variables, similar to the situation for   \textbf{annealed  disorder in Statistical Mechanics.}

Next we compute  an effective interaction  which is induced by the constraints. We integrate   the Lagrange multipliers $\varphi^{\alpha}(\vec{x},\widehat{T})$. We perform the integration for times $T>\frac{2\pi}{v_{F} \Lambda}$  where $\Lambda$ is the   momentum cut-off and  $ v_{F}$ is the Fermi velocity, which is maximum at half fillings and decreases with the doping. The product   $v_{F} \Lambda$ represents the electronic bandwidth in frequency units. Keeping only terms which are second order in  $\varphi^{\alpha}(\vec{x},\widehat{T})$ in the action,  $Log[\pi \frac{Sin(\varphi(\vec{x},\widehat{T}))}{\varphi(\vec{x},\widehat{T})}]$   allows us to integrate out the Lagrange multipliers and obtain the  effective interaction for times larger than  $T>\frac{2\pi}{v_{F} \Lambda}$   :

a) For electrons no interaction  is generated  and the physics is given by:

\begin{eqnarray}
S^{el.}[\widehat{t_{i}},\widehat{t_{f}}]&=&\int_{0}^{\pi} \frac{ds}{\pi}\int_{\widehat{t_{i}}}^{\widehat{t_{f}}}\,dt \int\frac{d\epsilon}{2\pi}\widehat{J}[\epsilon,s]\sum_{\sigma=\uparrow,\downarrow}[R^{\dagger (+)}_{\sigma}(t,\epsilon;s)(i\hbar \partial _{t}-\hbar\epsilon)R^{(+)}_{\sigma}(t,\epsilon;s)\nonumber \\& & +L^{\dagger (+)}_{\sigma}(t,\epsilon;s)(i\hbar \partial _{t}-\hbar\epsilon)L^{(+)}_{\sigma}(t,\epsilon;s) ] 
\end{eqnarray}

b) For holes we find the following  action is generated   at long times $T>\frac{2\pi}{v_{F} \Lambda}$:

\begin{eqnarray}
S^{eff-hole}[\widehat{t_{i}},\widehat{t_{f}}]&=&\int_{0}^{\pi} \frac{ds}{\pi}\int_{\widehat{t_{i}}}^{\widehat{t_{f}}}\,dt \int\frac{d\epsilon}{2\pi}\widehat{J}[\epsilon,s]\sum_{\sigma=\uparrow,\downarrow}[R^{\dagger (-)}_{\sigma}(t,\epsilon;s)(i\hbar \partial _{t}+\hbar\epsilon)R^{(-)}_{\sigma}(t,\epsilon;s) \nonumber \\&&+L^{\dagger (-)}_{\sigma}(t,\epsilon;s)(i\hbar \partial _{t}+\hbar\epsilon)L^{(-)}_{\sigma}(t,\epsilon;s) \nonumber \\ && + R^{\dagger (-)}_{\sigma}(t,\epsilon;s)\delta{\mu_{F}}(s)R^{(-)}_{\sigma}(t,\epsilon;s)+
L^{\dagger (-)}_{\sigma}(t,\epsilon;s)\delta{\mu_{F}}(s)L_{\sigma}(t,\epsilon;s)]\nonumber \\ && +\int_{0}^{\pi} \frac{ds_{1}}{\pi}\int_{0}^{\pi} \frac{ds_{2}}{\pi}\int_{\widehat{t_{i}}}^{\widehat{t_{f}}}\,dt_{1}\int_{\widehat{t_{i}}}^{\widehat{t_{f}}}\,dt_{2}\prod_{n=1}^4\frac{d\epsilon_{n}}{(2\pi)^3}\widehat{J}[\epsilon_{1},s_{1}]\widehat{J}[\epsilon_{2},s_{2}] \nonumber \\ &&g(s_{1}-s_{2};\widehat{T}) \delta(-\epsilon_{1}+\epsilon_{2}+\epsilon_{3}-\epsilon_{4}) \nonumber \\ &&[[ R^{\dagger (-)}_{\downarrow}(t_{1},-\epsilon_{1};s_{1}) L^{\dagger (-)}_{\uparrow}(t_{1},\epsilon_{2};s_{1})L^{(-)}_{\uparrow}(t_{2},\epsilon_{3};s_{2})R^{(-)}_{\downarrow}(t_{2},-\epsilon_{4};s_{2}) \nonumber \\ & & +L^{\dagger (-)}_{\downarrow}(t_{1},\epsilon_{1};s_{1}) R^{\dagger (-)}_{\uparrow}(t_{1},-\epsilon_{2};s_{1})L^{(-)}_{\uparrow}(t_{2},\epsilon_{3};s_{2})R{(-)}_{\downarrow}(t_{2},-\epsilon_{4};s_{2})  ]\nonumber \\& &+[ R^{\dagger (-)}_{\downarrow}(t_{1},-\epsilon_{1};s_{1}) L^{\dagger (-)}_{\uparrow}(t_{1},\epsilon_{2};s_{1})R^{(-)}_{\uparrow}(t_{2},-\epsilon_{3};s_{2})L^{(-)}_{\downarrow}(t_{2},\epsilon_{4};s_{2}) +\nonumber \\ & &L^{\dagger (-)}_{\downarrow}(t_{1},\epsilon_{1};s_{1}) R^{\dagger (-)}_{\uparrow}(t_{1},-\epsilon_{2};s_{1})L^{(-)}_{\uparrow}(t_{2},\epsilon_{3};s_{2})R^{(-)}_{\downarrow}(t_{2},-\epsilon_{4};s_{2})  ]]\nonumber \\ & &\equiv S^{hole}_{0}[\widehat{t_{i}},\widehat{t_{f}}]+S^{eff-hole}_{int.}[\widehat{t_{i}},\widehat{t_{f}}]
\end{eqnarray}

For a spherical Fermi surface  we have  $\widehat{J}[\epsilon,s]=1\equiv\epsilon^{0}$ and for all other cases we have 
 $\widehat{J}[\epsilon,s]\approx\epsilon^{z}$ where  $0\leq z \leq 1$,
where   $\Lambda$  is the momentum and energy $E=v_{F}\Lambda$ cut-off's.
This allows to  represent the coupling constant  by  $g(s_;\widehat{T})=\widehat{g}(s;\widehat{T})v_{F}E$.  We will use the $dimensionless$  $time$ $t=\frac{T v_{F}\Lambda}{2\pi}=\frac{TE}{2\pi}$  and  define a  real and imaginary coupling constant: $\widehat{g}(s;\widehat{T})\equiv-i(u-i\Delta)$  where the  $u\equiv u(t)=3(\frac{2\pi}{T v_{F}\Lambda})^2$ and  $\Delta $ is given by  $\Delta= u Sin(\delta)$ where  $\delta\rightarrow 0$.
The effective action $S^{eff-hole}_{int.}[\widehat{t_{i}},\widehat{t_{f}}]$ in eq.$(57)$ resemble the hole-hole interaction for Superconductivity at different times.
This form is obtained after we have replaced  the constraints operators  $Q^{(-)}_{\alpha}(\vec{x},t)$ by the chiral fermions  $R_{\sigma}(t,\epsilon;s)$  and $L_{\sigma}(t,\epsilon;s)$.

Next   we consider the situation away from half fillings,  without  the  umklapp terms.   Therefore we will not attempt to describe the Metal Insulator transition for which the presence of the exchange and umklapp interactions are   important. The action obtained is valid above a certain critical hole concentration $x>x_{c}$.
The  theory derived depends  on the bandwidth. The bandwidth decreases with the increases of the hole concentrations. Experimentally it is observed  that for large hole concentrations  the Fermi Liquid behavior is observed again.  At this stage we can not   prove  that we have a transition to a Fermi Liquid.  We can  show  that with increasing holes concentrations,  the  frequency region for which the self energy is linear in $\omega$ shrinks to zero.

\vspace{0.3 in}

\section{THE  SINGLE HOLE   GREEN'S FUNCTION}

\vspace{0.3 in}

The single particle excitations are the same as for the free electron case.
The only change is that the holes type excitations are  governed by the action given in equation $(57)$.

The definition of the  retarded Green's function for holes is given by: 
\begin{equation}
G^{(-)}[\vec{x},\sigma,t_{f};\vec{x'},\sigma',t_{i}]=i\vartheta[t_{i}-t_{f}]<\Psi^{0}|\mathbf{\psi}^{\dagger (-)}_{\sigma'}(\vec{x'},t_{i})\mathbf{\psi}^{(-)}_{\sigma}(\vec{x},t_{f})|\Psi^{0}>
\label{gren's}
\end{equation}

where $\vartheta[t_{i}-t_{f}]$ is the step function which is $1$  for time intervals  $t_{i}-t_{f}>0$ and $0$ otherwise.
An alternative representation for the Green's function is possible if we use the path integral representation \cite{Itz1}.
 In particular the  $matrix$ $elements$ of the  $evolution$  operator $U_{phys}[\overline{\psi}_{f},t_{f};\psi_{i},t_{i}]$   in the Grassmann  space \cite{Itz1} allows  us    to  compute   the Green's function. The relation between the path integral representation and the direct computation of the Green's function  $G^{(-)}[\vec{x},\sigma,t_{f};\vec{x'},\sigma',t_{i}]$  using  the ground state $|\Psi^{0}>$  has been shown in   \cite{Itz1,Greiner}.  In our case, the effect of the time dependent $coupling$ $constant$   can be investigated within the time dependent path integral given in the Grassmann space, as below:

\begin{eqnarray}
\mathcal{G}^{(-)}[\overline{\psi}_{i}(\vec{x'},\sigma',t_{i}),t_{f};\psi_{i}(\vec{x},\sigma,t_{f}),t_{f}]&=&i<\overline{\psi}_{i}(\vec{x'},\sigma',t_{i}),t_{i}|\mathbf{\psi}^{\dagger (-)}_{\sigma'}(\vec{x},t_{f})\mathbf{\psi}^{(-)}_{\sigma}(\vec{x},t_{f})|\psi_{f}(\vec{x},\sigma,t_{f}),t_{f}>\nonumber\\ & &  =i \int D(\overline{\psi}_{\sigma},\psi_{\sigma})
\overline{\psi}_{i}(\vec{x'},\sigma',t_{i})\psi_{f}(\vec{x},\sigma,t_{f})
 e^{S^{eff-hole}[t_{i},t_{f}]}
\end{eqnarray}

Following  \cite{Itz1,Greiner} we project the  $\mathcal{G}^{(-)}[\overline{\psi}_{f}(\vec{x'},\sigma',t_{f}),t_{f};\psi_{i}(\vec{x},\sigma,t_{i}),t_{i}]$ into the $ground$ $state$ $|\Psi^{0}>$
and obtain  the physical Green's function $G^{(-)}[\vec{x'},\sigma',t_{f};\vec{x},\sigma,t_{i}]$ defined in eq. $(58)$  which is represented  in  terms of the matrix elements in the Grassmann space.
\begin{eqnarray}
G^{(-)}[\vec{x'},\sigma',t_{f};\vec{x},\sigma,t_{i}]&= &\frac{<\overline{\psi}_{i}(\vec{x'},\sigma',t_{i})=0,t_{i}|\mathbf{\psi}^{\dagger (-)}_{\sigma'}(\vec{x},t_{i})\mathbf{\psi}^{(-)}_{\sigma}(\vec{x},t_{f})|\psi_{f}(\vec{x},\sigma,t_{f})=0,t_{f}>}{i<\overline{\psi}_{i}(\vec{x'},\sigma',t_{i})=0,t_{i}|\psi_{f}(\vec{x},\sigma,t_{f})=0,t_{f}>}\nonumber\\ & &=\frac{<F.S.|\mathbf{\psi}^{\dagger (-)}_{\sigma'}(\vec{x},t_{i})\mathbf{\psi}^{(-)}_{\sigma}(\vec{x},t_{f})T[e^{\frac{-i}{\hbar}S^{eff-hole}_{int.}[\widehat{t_{i}},\widehat{t_{f}}]}]|F.S.>}{i<F.S.|T[e^{\frac{-i}{\hbar}S^{eff-hole}_{int.}[\widehat{t_{i}},\widehat{t_{f}}]}]|F.S>}
\end{eqnarray}

where  $|F.S>=\prod_{\vec{K}=0}^{\vec{K}_{F}} \psi^{+}_{\uparrow}(\vec{K})\psi^{+}_{\downarrow}(\vec{K})|0>$ represents the Fermi Surface and $T \left[...\right]$ represents the time order.

The Physical Green's function   $G^{(-)}[T,\vec{K};\widehat{g}(\widehat{T}),T,\Lambda]$ 
 will be computed using the finite size action 
$S^{eff-hole}[\widehat{t_{i}},\widehat{t_{f}};\widehat{g}(\widehat{T}),T,\Lambda]$. The finite size effect is introduced by the duration of the hole excitations $t_{i}-t_{f}=T$.   The projection at times $t_{i}$ and $t_{f}$ generates a time dependent action.  
Due to the fact that the coupling constant $\widehat{g}(\widehat{T})$ is a function of the time duration for the hole excitations, it is advantageous to introduce a \textbf{parametric Green 's function $D^{(-)}[\vec{x'},\sigma',\tau_{f};\vec{x},\sigma,\tau_{i}]$} with the coupling constant $\widehat{g}(\widehat{T})$ and the  parametric time interval $\tau=\tau_{i}- \tau_{f}$ where,   $t_{f}\leq \tau_{f}\leq\tau_{i}\leq t_{i}$. 

\begin{eqnarray}
D^{(-)}[\vec{x'},\sigma',\tau_{f};\vec{x},\sigma,\tau_{i}]&= &\frac{<\overline{\psi}_{i}(\vec{x'},\sigma',t_{i})=0,t_{i}|\mathbf{\psi}^{\dagger (-)}_{\sigma'}(\vec{x},\tau_{i})\mathbf{\psi}^{(-)}_{\sigma}(\vec{x},\tau_{f})|\psi_{f}(\vec{x},\sigma,t_{f})=0,t_{f}>}{i<\overline{\psi}_{i}(\vec{x'},\sigma',t_{i})=0,t_{i}|\psi_{f}(\vec{x},\sigma,t_{f})=0,t_{f}>}\nonumber\\ & &=\frac{<F.S.|\mathbf{\psi}^{\dagger (-)}_{\sigma'}(\vec{x},\tau_{i})\mathbf{\psi}^{(-)}_{\sigma}(\vec{x},\tau_{f})T_{\tau}[e^{\frac{-i}{\hbar}S^{eff-hole}_{int.}[\widehat{t_{i}},\widehat{t_{f}}]}]|F.S.>}{i<F.S.|T_{\tau}[e^{\frac{-i}{\hbar}S^{eff-hole}_{int.}[\widehat{t_{i}},\widehat{t_{f}}]}]|F.S.>}
\end{eqnarray}

where $ T_{\tau}\left[... \right]$  stands for  the parametric  time   order. 

\vspace{0.1 in}

The computation of the single hole Green's  function will be done in two steps:

\vspace{0.1 in}

a) We compute first the   parametric  Green's function $D^{(-)} [\tau,\vec{K};\widehat{g}(\widehat{T}),T,\Lambda]$ where $\tau$ is the  correlation time interval and the coupling constant of the theory depends parametrically on the finite size $T$.
 For $\tau>0$  (holes excitations) the  Green's function $ D^{(-)} [\tau,\vec{K};\widehat{g}(\widehat{T}),T,\Lambda]$ is  computed with the help of the effective action $S^{eff-hole}_{int.}[\widehat{t_{i}},\widehat{t_{f}}]$  defined on the finite time interval $T$. 
Using the method of  finite size scaling \cite{Amit,Shahar} we will compute  the Green's function $D^{(-)} [\tau,\vec{K};\widehat{g}(\widehat{T}),T,\Lambda]$  and the Fourier transform 
$D^{(-)} [\omega,\vec{K};\widehat{g}(\widehat{T}),T,\Lambda]$  with respect the parametric time $\tau$ at  a fixed  temporal size $T$  and a  fixed  coupling constant $\widehat{g}(\widehat{T})$.

b) The physical Green's function $G^{(-)}[T,\vec{K};\widehat{g}(\widehat{T}),T,\Lambda]$  is related to the parametric Green's function  $ D^{(-)} [\tau,\vec{K};\widehat{g}(\widehat{T}),T,\Lambda]$:

 \vspace{0.1 in}

 $G^{(-)}[T,\vec{K};\widehat{g}(\widehat{T}),T,\Lambda]=\int_{0}^{\infty}\,d\tau\delta(\tau-T)D^{(-)} [\tau,\vec{K};\widehat{g}(\widehat{T}),T,\Lambda]$

\vspace{0.1 in}

 Once  the Green's function  $D^{(-)} [\omega,\vec{K};\widehat{g}(\widehat{T}),T,\Lambda]$  has been obtained,  the physical Green's function $G^{(-)}[\omega,\vec{K}]$  is evaluated using the Fourier transform properties:

\begin{equation}
G^{(-)}[\omega,\vec{K}]=\int_{\frac{2\pi}{E}}^{\infty}\,dT(\int_{-E }^{-\frac{2\pi}{|T|}}+\int_{\frac{2\pi}{|T|}}^{E})\,\frac{d\mathcal{\omega'}}{2\pi}
e^{i(\omega-\mathcal{\omega'})T}D^{(-)} [\omega',\vec{K};\widehat{g}(\widehat{T}),T,\Lambda]
\label{dreen's}
\end{equation}

Due to the finite size effect, the frequency integration $\omega'$ is restricted to $E>|\omega'|>\frac{2\pi}{T}$,  where $E$ is the  band width and $\frac{2\pi}{T}$ is the finite size frequency cut-off. The  bandwidth $E=v_{F}\Lambda$  is given by  $ \vec{v}_{F}=\frac{\vec{K}^{0}_{F}}{m}\sqrt{(1-x)}$ where  $\vec{K}^{0}_{F}$ is the Fermi momentum at half fillings and  $x$ represents the hole doping.  When  the    hole doping $x$ increases,   the bandwidth decreases $E(x\rightarrow 1)\rightarrow 0$.

 One of the interesting consequences of our formulation is that the single particle (hole) Green's function  is a function of the  effective time interval  $T$. The   coupling constant of the theory depends on  the time interval between  the creation and the  destruction of the hole.  Therefore we do not have one single action for all  the time intervals. For example, the single particle Green's function for an infinite time interval  $t_{i}-t_{f}\rightarrow \infty$ is described by the non interacting free action.   Using the  RG theory  we compute the  time dependent Green's function for a fixed  time interval   $t_{i}-t_{f}=T$.   
The frequency dependent  Green's function is rather non trivial since we have to perform a time integration over all the time and over all the possible coupling constants!
Performing the Fourier transform  by integrating over all the time dependent coupling constants (which are a function of the  time intervals $T$), we   will show   that the self energy  is dominated at low frequencies  by a relaxation  part which is linear in frequency $\Sigma_{Im}(\omega)\propto \omega F(\frac{\omega}{ v_{F}\Lambda})$.  The  function   $F(\frac{\omega}{v_{F}\Lambda})$  represents the crossover from $1$ when $\omega\rightarrow 0$  to $\omega$ for increasing frequencies.  The crossover region is determined by the bandwidth function   $v_{F}\Lambda$. With increasing doping the bandwidth decreases and the crossover  region shrinks to $0$.  As a result,    $\Sigma_{Im}(\omega)$ is modified to the Fermi liquid  behavior   $\Sigma_{Im}(\omega)\propto \omega^2$.

\vspace{0.3 in}

\section{THE  RENORMALIZATION  GROUP FOR THE  ACTION $S^{eff-hole}[\widehat{t_{i}},\widehat{t_{f}};\widehat{g}(\widehat{T}),T,\Lambda]$ IN TWO DIMENSIONS}

\vspace{0.3 in} 

The explicit form of  the  effective action in two space dimensions given  in eq.$(57)$  $S^{eff-hole}[\widehat{t_{i}},\widehat{t_{f}};\widehat{g}(\widehat{T}),T,\Lambda]$ is restricted by  the temporal size $T$. Therefore the method of  finite size scaling will be used.

The scaling dimensions for the coupling constant in eq.$(57)$ are  obtained from   a momentum cut-off  which is normal to the Fermi Surface \cite{Polchinski}. The scaling dimensions in the vicinity of  the Fermi Surface is  $d_{F.S.}=1$  (the scaling dimensions is only modified around the corners $(0,\pm\pi)$ , $(\pm\pi,0)$ close to half fillings). As a result, the  coupling constant  scales like  $g(\widehat{T})/v_{F}=\widehat{g}(\widehat{T})E^{2-d_{F.S.}}=\widehat{g}(\widehat{T})E$.
The fixed points for any theory are  achieved  by  taking  the limit  $b=e^l\rightarrow \infty $ where  $b=e^l$ describes the reduction of the bandwidth  cut-off
For the present problem  the limit $b=e^l\rightarrow \infty$ can not  be taken since we have to stop  the scaling at $b=b_{T}=v_{F}\Lambda T\equiv t $.

 The critical behavior $\widehat{g}(\widehat{T})E^{2-d_{F.S.}}$ is investigated using similar methods as  employed    for  the  $Ising- g\varphi^4$ model in  d=3 dimensions. For the Ising case the   coupling constant g obeys   $g=\widehat{g}E^{4-d}=\widehat{g}E$. (For the  $Ising- g\varphi^4$ one performs the calculations at a  $fictitious$ dimension $d=4-\eta$ such that at the value  $\eta=0$  the coupling constant is  marginal.  The R.G. equations take the form  $\frac{dg}{dl}=(4-d)g...=\eta g...$;  to recover the physics for  d=3  we take the limit $\eta\rightarrow 1$ at the end of the calculation.)

Following  the analogy with the $Ising- g\varphi^4$ R.G. we introduce  fictitious dimensions of the Fermi Surface $d_{F.S.}=2-\eta$, such that when $\eta\rightarrow 1$  one reproduces the  one dimensional scaling of the Fermi Surface.
The integration variable    $\frac{d\epsilon}{(2\pi)}\widehat{J}[\epsilon,s]$ is replaced by $\frac{d^{2-\eta}\epsilon}{(2\pi)^{2-\eta}}\widehat{J}_{2-\eta}[\epsilon,s]$ such that 
at the limit $\eta\rightarrow 1$  we obtain $\widehat{J}_{2-\eta}[\epsilon,s]\rightarrow\widehat{J}[\epsilon,s]$. 
 As a result, the scaling dimension of the coupling constant becomes marginal for $\eta=0$.
For $ 0<\eta\leq1$ we find:

\begin{equation}
g(\widehat{T})=\widehat{g}(\widehat{T})E^{2-(2-\eta)}=\widehat{g}(\widehat{T})E^{\eta}
\label{eta}
\end{equation}

We will use \textbf{the differential   R.G. method}  where the integration in the energy   shell $E-dE\leq \epsilon \leq E$ is performed using   the differential variable  $dl=\frac{dE}{E}$  \cite{Schm,Shankar,Kopietz}. We find  that the coupling constant   constant  $\widehat{g}(\widehat{T})$ for  the $zero$ $angular$ $momentum$ $channel$ \cite{Shankar} obeys the following R.G. equation.
   
\begin{equation}
\frac{d\widehat{g}}{dl}=\eta\widehat{g}-\frac{(\widehat{g})^2}{4\pi}\hat{\beta}(t)
\label{rG}
\end{equation}

Comparing eq.$(64)$ with the $R.G.$ equation for singlet superconductivity \cite{Shankar},
we observe that due to the additional time integration in eq.$(57)$, eq.$(64)$ has a linear term  $\widehat{g}$ with the scaling dimension $\eta=1$ and  that the term $\widehat{g}^2$ is rescaled  by the temporal finite size parameter $\hat{\beta}(TE=t)$.  The dimensionless parameter obeys $1<\hat{\beta}(TE)<|T|E= |T| v_{F}\Lambda \equiv t$. At the limit $T\rightarrow  \infty$  one  finds that the R.G. equation has an   $infrared$ $stable$ $fixed$ $point$   $\widehat{g}=\frac{4\pi}{\eta \hat{\beta}(TE)}\rightarrow 0$,  which describes the  Marginal Fermi liquid. 
We use the  complex  representation  for  the coupling constant
$ \widehat{g}= -i(u+i\Delta)$  with $u=3(\frac{1}{|T| v_{F}\Lambda})^2\equiv\frac{ 3}{t^2}$
and $\Delta= u Sin(\delta)$, with  the initial condition  $\delta\rightarrow 0$. We find that the equations have  an  infrared  stable  fixed point given  by $(u^*\rightarrow 0,\Delta^*=\frac{4\pi}{\eta \hat{\beta}(TE\equiv t)}\rightarrow 0)$. Due to the finite temporal size $T$  the R.G. equation $(64)$ is  only valid  for $ 0<l<Log(|T| v_{F}\Lambda)\equiv Log(t)\equiv l_{t}$.

In order to construct the full R.G.  flow, we have to compute the differential self energy from which we will extract the wave function renormalization.
We will work with the two dimensional representations of the coupling constant $\widehat{g}=-i u+\Delta$,  with the initial conditions $\Delta(l=0)\rightarrow 0$.
The self energy  of our action is a function of frequency and coupling constants  $\Sigma[\omega,\epsilon;u,\Delta,E]$.   We will use  the infrared  stable fixed point to compute the  self energy  $\Sigma[\omega,\epsilon;u,\Delta,E]$  and to  tune  the chemical potential $\delta{\mu_{F}}$.  We find that $\delta{\mu_{F}}(s)$ is given by the same  self energy at zero frequency  for all the points $s$ on the Fermi surface, $\delta{\mu_{F}}(s)=\Sigma[0,0;u^*,\Delta^*]$. Expanding the self energy in powers of  $\omega$ allows to compute the $wave$ $function$ $renormalization$ $Z_{\psi}$.  We find: 
$\frac{d Log[Z_{\psi}(u(l)]}{dl}=\frac{u(l)}{2\pi}$.

 As a result,  the previous R.G. equations for  $u$ and $\Delta$ are modified:   

\begin{eqnarray}
\frac{du}{dl}&=&(\eta -\frac{u}{\pi})u-\hat{\beta}(t)\frac{u \Delta}{2\pi}\nonumber\\ 
\frac{d\Delta}{dl}&=&(\eta-\frac{u}{\pi})\Delta -\hat{\beta}(t)\frac{\Delta^2-u^2}{4\pi}
\end{eqnarray}

This set of equations have  the initial conditions
 $u(l=0)=3(\frac{1}{T v_{F}\lambda})^2\equiv \frac{3}{t^2}$
and $\Delta(l=0)= u(l=0) Sin(\delta)\rightarrow 0$. 
Due to the finite time interval $T$  we have to $restrict$ the scaling  to the domain  $1<\hat{\beta}(TE\equiv t)< T v_{F}\Lambda\equiv t$. 

As a result, the  coupling constant will reach  the end point values $u(l_{t})=u(Log(t))$ and $\Delta(l_{t})=\Delta(Log(t))$.
The finite size scaling results are given by  the numerical solution of the R.G equations.

This set of equations have an $infrared$ $stable$ $fixed$ $point$      given by $(u^{*}=0,\Delta^{*}=0)$.  
For a finite time interval $T$ the values of the coupling constant deviate from the fixed point. For $t\rightarrow \infty $  we obtain  that the coupling constants reach the values  $u(l_{t})=\frac{u_{1}}{t}$     and  $\Delta(l_{t})=\frac{d_{1}}{t}$, where $u_{1}$ and $d_{1}$   are universal constants. 
The result of the R.G. flow are given in figure 1. In figure 1  we show that the $dissipative$ $coupling$ $constant$  $\Delta(l_{t})$  fits the analytic form $d(t)=0.7588/t $ with  the universal constant  $d_{1}=0.7588$.

\vspace{0.2 in}

 \begin{figure}
\includegraphics[width=3.5 in ]{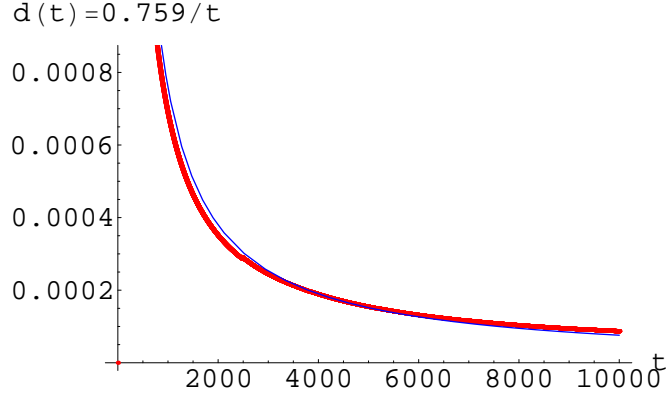}
\caption{(Color online)The Dissipative Coupling  Constant $\Delta(l_{t})$ : We show two graphs one is the R.G. result and the second is the fit to the analytic   form $d(t)=0.7588/t \equiv d_{1}/t$ }
\end{figure}

\vspace{0.3 in}

\section{COMPUTATION OF THE  PARAMETRIC   GREEN'S  FUNCTION $D$ FOR  TIMES   $\tau\leq T$}

\vspace{0.3 in}

Using the R.G. results from the previous section we will compute the hole type  parametric Greens function  for parallel  spins $D^{(-)} [\omega,\epsilon; \widehat{g}= -i(u+i\Delta)]\equiv D^{(-)} [\omega,\epsilon;u,\Delta]$, where $\epsilon$ is the energy excitation  perpendicular to the Fermi Surface $\epsilon=\vec{v_{F}}\cdot(\vec{K}-\vec{K_{F}}(s))$.

This Greens function $D^{(-)} [\omega,\epsilon;u,\Delta]$ is computed using the unperturbed Green's function $D^{(-)}_{0}[\omega,\epsilon]$ 
 
\begin{eqnarray}
(D^{(-)} [\omega,\epsilon;u,\Delta,E])^{-1}= (D^{(-)}_{0}[\omega,\epsilon])^{-1}+\Sigma[\omega,\epsilon;u,\Delta,E]\nonumber\\= \omega-\epsilon-\delta{\mu_{F}}+\Sigma[0,0;u,\Delta,E]+\omega\frac{d\Sigma[\omega,\epsilon;u,\Delta,E]}{d\omega}
\end {eqnarray}

where $\Sigma[\omega,\epsilon;u,\Delta,E]$ is the self energy. Using the fact that the R.G. equation has  an infrared fixed point we  can tune the shift in the chemical potential $\delta{\mu_{F}}(s)$ such that $\delta{\mu_{F}}(s)=\Sigma[0,0;u(l_{t})\approx u^* \approx  \frac{u_{1}}{t}  ,\Delta(l_{t})\approx \Delta^*\approx \frac{d_{1}}{t},E]$.

We obtain the wave function renormalization $Z_{\psi}(u(l_{t}))=e^{-\int_{0}^{_{t}}u(l')\,dl'}$.  In the present case we stop scaling at   $e^{l}\equiv v_{F}\Lambda T\equiv t$ and  we find:

\begin{equation}
Z_{\psi}[u(l=log(t)=l_{t})]=e^{-\int_{0}^{l=log(t)=l_{t})}u(l')\,dl'}\approx e^{-\frac{u_{1}}{t}}
\label{zwave}
\end{equation}

As a result we obtain  the  finite size  Green's function for $t>>1$ with  the  universal parameters $u_{1}$ and $d_{1}$:
$D^{(-)} [\omega,\epsilon;u(l_{t}),\Delta(l_{t}),E]=\frac{e^{-u(l_{t})}}{\omega-\epsilon +i\omega\frac{\Delta(l_{t})}{2\pi}} \approx\frac{e^{-\frac{u_{1}}{t}}}{\omega-\epsilon +i\omega\frac{d_{1}}{2\pi t}}$. The action in eq.$(57)$ is restricted for large  time intervals  $T>\frac{2\pi}{v_{F}\Lambda}$ for which we can replace  $e^{-\frac{u_{1}}{t}}\rightarrow 1$.

The parametric Green's function which is restricted to the frequency interval $\frac{2\pi}{T}<\omega< v_{F}\Lambda$  is given by: 

\begin{equation}
D [\omega,\epsilon;u(l_{t}),\Delta(l_{t}),E]=\frac{\vartheta[\epsilon]}{\omega-\epsilon+i\delta}+ \frac{ \vartheta[-\epsilon]}{\omega-\epsilon +i\omega\frac{d_{1}}{2\pi t}}
\label{fullG}
\end{equation}

\vspace{0.3 in}

\section{COMPUTATION OF THE  PHYSICAL   GREEN'S  FUNCTION $G$ } 

\vspace{0.3 in}

The Physical Green's function $G$ will be computed  from the Fourier representation   of the parametric  Green's function $D$ given by  equation  $(68)$.  We substitute in equation $(62)$ the explicit form of the parametric Green's function  $D[w,\epsilon]=D_{Re}[w,\epsilon]+iD_{Im}[w,\epsilon]$ as given in equation $(68)$ .

Using the Fourier transform of equation $62$ we find that the Green's function $G$ is given in terms of two $new$ $functions$ $F_{R}(t;\omega,\epsilon)$ and $F_{Im}(t;\omega,\epsilon)$ which are a linear combination of  the real and imaginary part of the parametric Green's function $D^{(-)}[w,\epsilon]=D^{(-)}_{Re}[w,\epsilon]+iD^{(-)}_{Im}[w,\epsilon]$.

\vspace{0.2 in}

$G^{(-)}_{Re}[w,\epsilon]=\int_\frac{2\pi}{v_{F}\Lambda}^T F_{R}(t;\omega,\epsilon)\,dt$

\vspace{0.2 in}

$G^{(-)}_{Im}[w,\epsilon]=\int_\frac{2\pi}{v_{F}\Lambda}^{T}F_{Im}(t;\omega,\epsilon)\,dt$

\vspace{0.2 in}

where 

\begin{eqnarray}
F_{R}(t;\omega,\epsilon)&=&\int_\frac{2\pi}{T}^{v_{F}\Lambda}\frac{d\varpi}{2\pi}[D^{(-)}_{Re}[\varpi,\epsilon]Cos((\omega-\varpi)t)-D^{(-)}_{Re}[-\varpi,\epsilon]Cos((\omega+\varpi)t)]\nonumber\\ & &-\int_\frac{2\pi}{T}^{v_{F}\Lambda}\frac{d\varpi}{2\pi}[D^{(-)}_{Im}[\varpi,\epsilon]Sin((\omega-\varpi)t)-D^{(-)}_{Im}[-\varpi,\epsilon]Sin((\omega+\varpi)t)]
\end{eqnarray}

\begin{eqnarray}
F_{Im}(t;\omega,\epsilon)&=&\int_\frac{2\pi}{T}^{v_{F}\Lambda}\frac{d\varpi}{2\pi}[D^{(-)}_{Re}[\varpi,\epsilon]Sin((\omega-\varpi)t)-D^{(-)}_{Re}[-\varpi,\epsilon]Sin((\omega+\varpi)t)]\nonumber\\ & &-\int_\frac{2\pi}{T}^{v_{F}\Lambda}\frac{d\varpi}{2\pi}[D^{(-)}_{Im}[\varpi,\epsilon]Cos((\omega-\varpi)t)-D^{(-)}_{Im}[-\varpi,\epsilon]Cos((\omega+\varpi)t)]
\end{eqnarray}

The Physical Green's function  $G [\omega,\epsilon=\vec{v_{F}}\cdot(\vec{K}-\vec{K_{F}}(s));E=v_{F}\Lambda]$ is given in terms of the  self energy $\Sigma(\omega,\epsilon)=\Sigma_{Re}(\omega,\epsilon)+i \Sigma_{Im}(\omega,\epsilon)$.

\begin{equation}
G [\omega,\epsilon;E]=\frac{\vartheta[\epsilon]}{\omega-\epsilon+i\delta}+ \frac{ \vartheta[-\epsilon]}{\omega-\epsilon + \Sigma_{Re}(\omega)+i \Sigma_{Im}(\omega)}
\label{finaq}
\end{equation}

In the present case equations $(69)$ and $(70)$ gives us   the Physical Green's in  terms of the parametric   Green's  function $D$.  We represent the self energies of $G$  in terms of the parametric Green's function.
We find that:

 $\Sigma_{Re}(\omega,\epsilon;z)= G_{Re}[w,\epsilon;z]/((G_{Re}[w,\epsilon;z])^2+(G_{Im}[w,\epsilon;z])^2) -w$
 
 and 
 
 $\Sigma_{Im}(\omega,\epsilon;z)= -G_{Im}[w,\epsilon;z]/((G_{Re}[w,\epsilon;z])^2+(G_{Im}[w,\epsilon;z])^2) -z$  where  $z\rightarrow 0$
 
 where   $G_{Re}[w,\epsilon]$ and  $ G_{Im}[w,\epsilon]$ are given by the  equations $(69)-(70)$ .

 The results for the self energy   are given in figures 2 and 3 at a fixed energy  $\epsilon=0$. (The  Green's function will be  given as a function of dimensionless frequency  and energy $ \frac{\omega}{E=v_{F}\Lambda}\rightarrow \omega$  and  $ \frac{\epsilon}{E=v_{F}\Lambda}\rightarrow \epsilon$.)
  We observe that with increasing frequency the calculation of the self energy at energy $\epsilon=0$  becomes less accurate.  For larger frequencies we have to compute $\Sigma_{Re}(\omega,\epsilon) $ 
 and $\Sigma_{Im}(\omega,\epsilon)$  at finite energies $\epsilon$.
 
 The self energy $\Sigma_{Im}(\omega)\propto \omega F(\frac{\omega}{ v_{F}\Lambda})$  is  linear in frequency for $\omega\rightarrow 0$.  The  function   $F(\frac{\omega}{v_{F}\Lambda})$  represents the crossover from $1$ when $\omega\rightarrow 0$  to $\omega$ for increasing frequencies. When doping  increases,   the bandwidth   $v_{F}\Lambda$ decreases and  we observe that the linear region  in frequency shrinks to $0$.  As a result    $\Sigma_{Im}(\omega)$ is modified to the Fermi liquid  behavior   $\Sigma_{Im}(\omega)\propto \omega^2$.

 \begin{figure}
\includegraphics[width=3.5 in ]{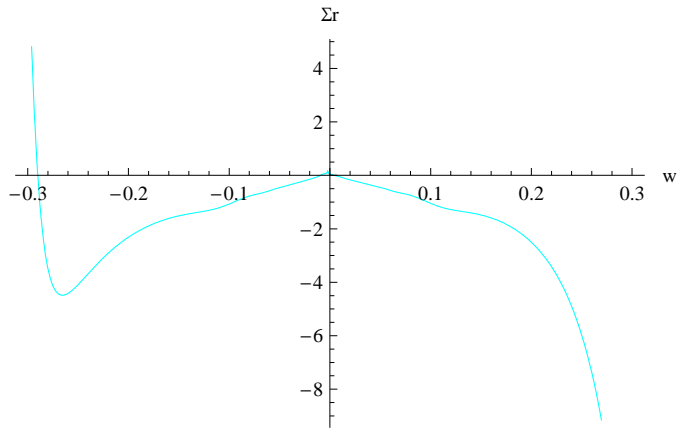}
\caption{(Color online)The real part  of the self energy self energy  of the physical Green's function $G(\omega,\epsilon)$ computed from the finite  size Green's function $D(\omega,\epsilon; u(l_{t}),\Delta(l_{t}),E)$}
\end{figure}

\begin{figure}
\includegraphics[width=3.5 in ]{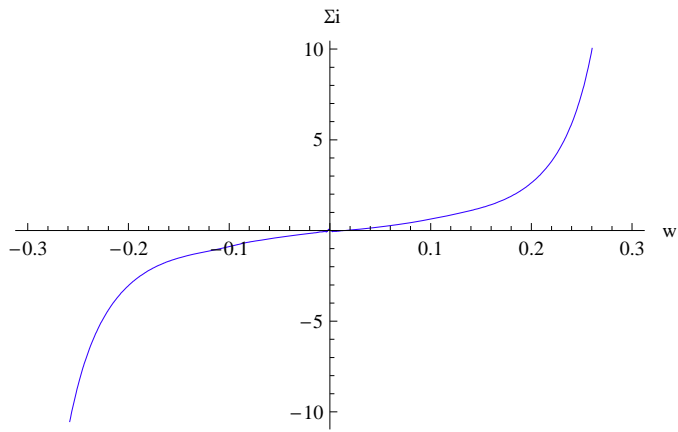}
\caption{(Color online)The Imaginary  part of the self energy of the physical Green's function $G(\omega,\epsilon)$ computed from the finite  size Green's function $D(\omega,\epsilon; u(l_{t}),\Delta(l_{t}),E)$}
\end{figure}

 Figure $3$ shows that the imaginary part of the self energy  for low holes densities  is linear in  frequency.  As a result,  the single hole excitation  (at low  frequencies) has  a width which is  linear in frequency, and the  scattering rate obeys  $\frac{1}{\tau}\propto \omega F(\frac{\omega}{v_{F}\Lambda})
$ in agreement with  the infrared data \cite{Schlesinger}. 
 
\vspace{0.3 in}

\section{APPLICATION   OF THE THEORY TO SUPERCONDUCTIVITY}

\vspace{0.3 in}

In this section we will attempt to connect the theory presented  with the physics of the high $T_{c}$ material. In particular we have considered a model at  $zero$ $temperature$ away from half fillings   where  the magnetic order has $vanished$.
Under this condition we have shown that the marginal Fermi liquid with an imaginary self energy which is linear in frequency is obtained.
This results are in agreement with the   experiments which  show that at T=0  a window exists between the magnetic ordered state and the appearance of superconductivity.
Therefore, when the   exchange interaction is included  a d-wave  superconducting  \cite{hdavid} phase within the exclusion of double occupancy will appear.  The $d-wave$ order   parameter  $\sum_{\vec{a}}[\mathbf{\psi}_{\uparrow}(\vec{x})\mathbf{\psi}_{\downarrow}(\vec{x}+\vec{a})-\mathbf{\psi}_{\downarrow}(\vec{x})\mathbf{\psi}_{\uparrow}(\vec{x}+\vec{a})]|\Psi^{0}>\neq 0$ ($contrarily$ to the $s-wave$)  is   compatible with constraints $\mathbf{\psi}_{\uparrow}(\vec{x})\mathbf{\psi}_{\downarrow}(\vec{x})|\Psi^{0}>=0$.  Further doping of the superconductor at T=0 will give rise to a transition from a free vortex monopole phase to a spin wave phase \cite{FSch}.

For the remaining part of this section we will show that the effect of exclusion of double occupancy gives rise in the superconducting phase to an asymmetry in the  tunneling density of states.
In order to demonstrate   the asymmetry effect of the projection of double occupancy  we consider a qualitative  calculation for   superconductors . Strictly speaking  an  accurate comparison with the experiment must use the  full  $d$ wave structure.  In order  to demonstrate the effect of asymmetry induced by the projection, it is important  to show that the   asymmetry can be obtained also for an  uniform  state. 
We consider the standard $BCS$ hamiltonian and use the projection introduced in the previous sections.
In the absence of the projection the effect of the $BCS$  gap $\Delta_{BCS}$ gives rise (after integration over the single particle energy) to the following  $tunneling$ $density$ of $states$,  $N_{T}^{(S-wave)}(\omega)=\frac{2}{\pi}\frac{K_{F}}{v_{F}}(\frac{-1}{2\pi})\int d\epsilon Im.G^{0-S}(\epsilon,\omega)$: 

\begin{eqnarray}
N_{T}^{(S-wave)}(\omega)=\frac{2}{\pi}\frac{K_{F}}{v_{F}}\frac{1}{\pi}\int_{0}^{1}\,d\epsilon [(1+\frac{\epsilon}{\sqrt{\epsilon^2 +\Delta_{BCS}^2}})(\frac{z}{(\omega-\sqrt{\epsilon^2 +\Delta_{BCS}^2})^2 +z^2}) \nonumber\\ +(1+\frac{\epsilon}{\sqrt{\epsilon^2 +\Delta_{BCS}^2}})(\frac{z}{(\omega+\sqrt{\epsilon^2 +\Delta_{BCS}^2})^2 +z^2})]
\end{eqnarray}

where  $z\rightarrow 0$. 
 
\vspace{0.2 in}

Next we repeat the calculation when we project out double occupancy!
For this purpose we use  the $self$ $energy$ given in figure $3$,  $\Sigma_{Im}(\omega)$. 

The tunneling density of states for the $ projected$  case  of  double occupancy is  given by
$N_{T}^{(S-w.ex.)}(\omega)$:

\vspace{0.2 in}

\begin{eqnarray}
N_{T}^{(S-w.ex.)}(\omega)=\frac{2}{\pi}\frac{K_{F}}{v_{F}}\frac{1}{\pi}\int_{0}^{1}\,d\epsilon [(1+\frac{\epsilon}{\sqrt{\epsilon^2 +\Delta_{BCS}^2}})(\frac{z}{(\omega-\sqrt{\epsilon^2 +\Delta_{BCS}^2})^2 +z^2}) \nonumber\\ +(1+\frac{\epsilon}{\sqrt{\epsilon^2 +\Delta_{BCS}^2}})(\frac{\Sigma_{Im}(\omega)}{(\omega+\sqrt{\epsilon^2 +\Delta_{BCS}^2})^2 +(\Sigma_{Im}(\omega))^2})]
\end{eqnarray}

where  $z\rightarrow 0$. 

In order to emphasize  the asymmetric effect of the self energy we consider typical values of temperatures and gap.  For the gap we take the value   $\Delta_{BCS}=0.38 \times 10^{-3}$ eV and  restrict the temperature  to, $T_{B}<\Delta_{BCS}$.

In figure $4$ we show the two graphs of the tunneling density of states, $N_{T}^{(S-wave)}(\omega=\frac{eV}{\hbar})$ is the tunneling density of states  in the absence of projection and $N_{T}^{(S-w.ex.)}(\omega=\frac{eV}{\hbar})$ is the tunneling density of states for the projected case. The tunneling density of states as a function  of the tunneling voltage  $V$ shows   a clear asymmetry between the projected $N_{T}^{(S-w.ex.)}$ and the non - projected  function $N_{T}^{(S-wave)}$.

\begin{figure}
\includegraphics[width=3.5 in ]{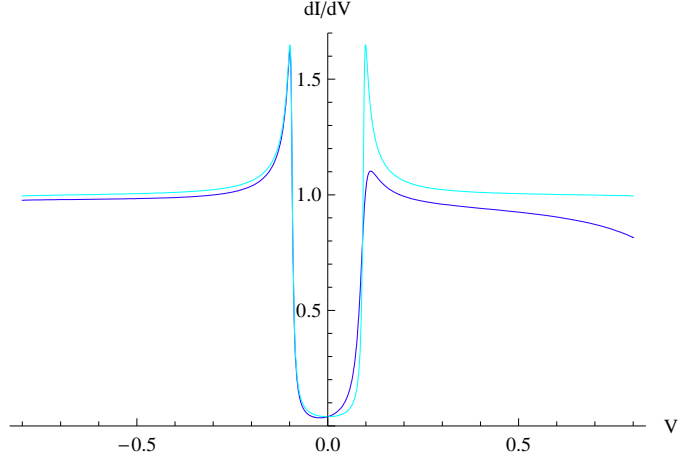}
\caption{(Color online)a)-The tunneling density of states  for S-wave superconductivity as a function of the voltage $V$
 without projection $N_{T}^{(S-wave)}(\frac{eV}{\hbar})$ 
  b)-The tunneling density of states  for S-wave superconductivity as a function of the voltage $V$
with   projection  (the $asymmetric$ graph) $N_{T}^{(S-w.ex.)}(\frac{eV}{\hbar})$    }
\end{figure}
 
In figure $5$ we show on the same graph: the experimental data for the tunneling of the density  states  observed in  \cite{Pan} and our projected tunneling density of states.
 Figure $5$ demonstrates that the asymmetry in the tunneling density of states can be explained by the projected Green's function.

 \begin{figure}
\includegraphics[width=3.5 in ]{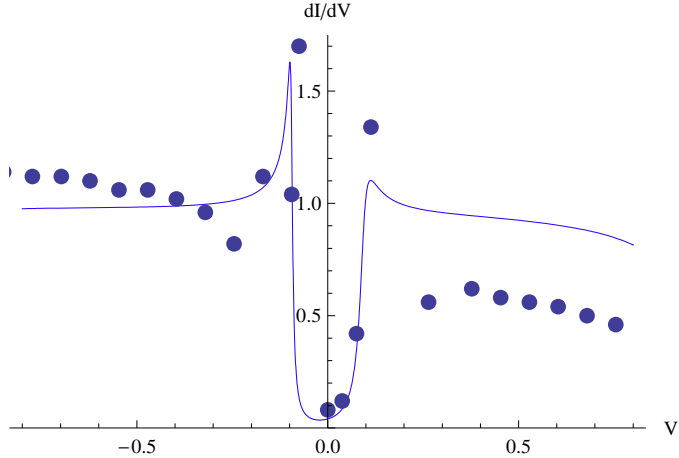}
\caption{(Color online)The tunneling density of states  given by the dots extracted from the experiment in ref. \cite{Pan} compared to the analytic  formula  for the projected tunneling density of states  $N_{T}^{(S-w.ex.)}(\frac{eV}{\hbar})$}
\end{figure}

 \vspace{0.3 in}

\section{CONCLUSION}

 \vspace{0.3 in}

We have proposed a new method to  study   correlated electrons where the  traditional method of slave particles is avoided.
We  have demonstrated that by  choosing  proper variables additional constraints can be  included. We obtain a system of    first class constraints which   generate gauge transformations.  

 The authors   \cite{Muthu,Gutz} have computed  the wave function  using  only one constraint. We identify   two  additional constraints  which form an non-Abelian group and neglect according to the R.G. analysis  the secondary constraints. As a result   the simple delta function   constraint is replaced by  a non-linear  integration measure   which  gives rise to time dependent action.
 
 The effective action is   analyzed with the help  of the R.G.method   for finite size systems (in the time domain).
 
The R.G. analysis  allows to compute the single particle self energy which is used for $qualitative$ comparison with the experiments.

\vspace{0.8 in}

\textbf{APPENDIX:THE SECONDARY FIRST CLASS  CONSTRAINTS}

\vspace{0.4 in}

In this Appendix we   will show that the effective interactions induced by the secondary first class  constraints  are irrelevant operators  for describing the  long low energy
Physics  and therefore can be neglected.

In order to show this we perform the followings steps:

\vspace{0.1 in}

A1)  Compute the commutator of the kinetic energy with the primary first class constraints. (In the absence of the  exchange interaction, the hamiltonian is given by  $H_{0}$.) 

\vspace{0.1 in}

$[\mathbf{\psi}_{\sigma=\downarrow}(\vec{x})\mathbf{\psi}_{\sigma=\uparrow}(\vec{x}),H_{0}]=t\sum_{\vec{a}}[\mathbf{\psi}_{\sigma=\downarrow}(\vec{x})\mathbf{\psi}_{\sigma=\uparrow}(\vec{x}+\vec{a})+\mathbf{\psi}_{\sigma=\downarrow}((\vec{x}+\vec{a}))\mathbf{\psi}_{\sigma=\uparrow}(\vec{x})]$
 
The  secondary  constraints  $q(\vec{x})$, $q^{+}(\vec{x})$ and $q_{3}(\vec{x})$ are obtained by commuting the primary constraints with the hamiltonian $H_{0}$ and subtracting the primary constraints: 

\vspace{0.1 in}

$q(\vec{x})\equiv \frac{1}{t}[Q(\vec{x}),H_{0}]-Q(\vec{x})= \sum_{\vec{a}}[\mathbf{\psi}_{\sigma=\downarrow}(\vec{x})\mathbf{\psi}_{\sigma=\uparrow}(\vec{x}+\vec{a})+\mathbf{\psi}_{\sigma=\downarrow}((\vec{x}+\vec{a}))\mathbf{\psi}_{\sigma=\uparrow}(\vec{x})] -Q(\vec{x})$

\vspace{0.1 in}

$q^{+}(\vec{x})\equiv \frac{1}{t}[Q^{+}(\vec{x}),H_{0}]-Q^{+}(\vec{x})= \sum_{\vec{a}}[\mathbf{\psi}_{\sigma=\downarrow}(\vec{x})\mathbf{\psi}_{\sigma=\uparrow}(\vec{x}+\vec{a})+\mathbf{\psi}_{\sigma=\downarrow}((\vec{x}+\vec{a}))\mathbf{\psi}_{\sigma=\uparrow}(\vec{x})]^{+} -Q^{+}(\vec{x})$

and 
\vspace{0.1 in}

$q_{3}(\vec{x})\equiv \frac{1}{t}[Q_{3}(\vec{x}),H_{0}]-[Q_{3}(\vec{x})-1)]= \sum_{\vec{a}}[\mathbf{\psi}_{\sigma=\downarrow}(\vec{x})\mathbf{\psi}_{\sigma=\downarrow}(\vec{x}+\vec{a})+\mathbf{\psi}_{\sigma=\uparrow}((\vec{x}+\vec{a}))\mathbf{\psi}_{\sigma=\uparrow}(\vec{x})]  -[Q_{3}(\vec{x})-1]$

\vspace{0.1 in}

A2) In  section VII we have  parametrized Fermi-Surface in terms of the polar angle $s$, normal  $\hat{N}(s)$  to  the Fermi Surface and chiral fermions $R_{\sigma}(\vec{x},s)$, ,$L_{\sigma}(\vec{x},s)$ .The hamiltonian  $H_{0}$  is given by: 

\vspace{0.1 in}

$H_{0}=\int d^{2}x
\int_{0}^{\pi} \frac{ds}{\pi} \sum_{\sigma=\uparrow,\downarrow}[R^{\dagger}_{\sigma}(\vec{x};s)(-i v_{F}\hbar)\hat{N}(s)\cdot\vec{\partial}_{\vec{x}}R_{\sigma}(\vec{x};s)+L^{\dagger}_{\sigma}(\vec{x};s)(i v_{F}\hbar)\hat{N}(s)\cdot\vec{\partial}_{\vec{x}}L_{\sigma}(\vec{x};s)]$ 

A3) Approximating the difference between the commutators  and the primary constraints by a first order spatial derivative around the Fermi-Surface  we obtain:

\vspace{0.1 in}

$q(\vec{x})\approx \int_{0}^{\pi}\frac{ds}{\pi}[R_{\uparrow}(\vec{x},s)\hat{N}(s)\cdot\vec{\partial}_{\vec{x}}L_{\downarrow}(\vec{x},s)-L{\downarrow}(\vec{x},s)\hat{N}(s)\cdot\vec{\partial}_{\vec{x}}R_{\downarrow}(\vec{x},s)]$

\vspace{0.1 in}

$q^+(\vec{x})\approx[\int_{0}^{\pi}\frac{ds}{\pi}[R_{\uparrow}(\vec{x},s)\hat{N}(s)\cdot\vec{\partial}_{\vec{x}}L_{\downarrow}(\vec{x},s)-L{\downarrow}(\vec{x},s)\hat{N}(s)\cdot\vec{\partial}_{\vec{x}}R_{\downarrow}(\vec{x},s)]]^{+}$ 

and 
\vspace{0.1 in} 

$q_{3}(\vec{x})\approx \int_{0}^{\pi}\frac{ds}{\pi}[R^{+}_{\uparrow}(\vec{x},s)\hat{N}(s)\cdot\vec{\partial}_{\vec{x}}R_{\uparrow}(\vec{x},s)+L^{+}{\downarrow}(\vec{x},s)\hat{N}(s)\cdot\vec{\partial}_{\vec{x}}L_{\downarrow}(\vec{x},s)]$.

A4) The  presence of the  secondary constraints  the modifies  the  Lagrangian $L\Rightarrow L+\delta L$ where $\delta L$ is given by:

$\delta L =\sum_{r=1}^{3}\hat{\pi}_{r}(\vec{x},t)\partial_{t}\hat{\lambda}^{r}(\vec{x},t)-\sum_{r=1}^{3}(\hat{\pi}_{r}(\vec{x},t)\hat{\chi}^{r}(\vec{x},t)+\hat{\lambda}^{r}(\vec{x},t)q^{(-)}_{r}(\vec{x},t))  $
    
where  $\hat{\lambda}^{r}(\vec{x},t)$   are the Lagrange multipliers which enforces the secondary  constraints, $\hat{\pi}_{r}(\vec{x},t)$ are  the canonical momentum conjugate to the new Lagrange  $\hat{\lambda}^{r}(\vec{x},t)$ multipliers and $q^{(-)}_{r}(\vec{x},t))$ are obtained using  the linear transformation given in equations $(35)$ and $(36)$. This gives rise to the evolution operator:

$\hat{U}_{phys}[t_{f},t_{i}]=e^{\frac{-i}{\hbar}(t_{f}-t_{i})H}
\int \prod_{\alpha=1}^{3}\mathcal{D}\lambda_{\alpha}(T)\prod_{r=1}^{3}\mathcal{D}\hat{\lambda}_{r}(T) e^{\frac{-i}{{\hbar}}\left[ \sum_{\alpha=1}^{3)}\int\lambda_{\alpha}(\vec{x})Q^{-}_{\alpha}(\vec{x})\,d^{d}x+\sum_{r=1}^{3)}\int\hat{\lambda}_{r}(\vec{x})q^{(-)}_{r}(\vec{x}))\,d^{d}x \right]  }$

\vspace{0.1 in}

In order to compute  the effective interaction induced by   the secondary constraints  we need  the integration measure for the  secondary Lagrange multipliers  $\hat{\lambda}^{r}(\vec{x},t)$.   We approximate the measure by a regular integration and   obtain a set of   delta functions which enforces  the constraints $q^{(-)}_{r}(\vec{x},t))$.   The delta functions constraints   effectively replaced by  exponentials  of   Gaussian terms  with the coupling constant $\delta g_{I}$ ( which at short distances goes to infinity   and therefore is equivalent to a delta function). The Gaussian  action is given by; $\delta g_{I}\sum_{r=1}^{3}\int d^{d} x \int dt_{1}\int dt_{2}[q^{(-)}_{r}(\vec{x},t_{1})) q^{(-)}_{r}(\vec{x},t_{2}))]$.  As a result, we obtain the correction  $\delta S^{eff-hole}_{int.}[\widehat{t_{i}},\widehat{t_{f}}]$  to the original action given in eq.$(57)$:
\begin{eqnarray*}
\delta S^{eff-hole}_{int.}[\widehat{t_{i}},\widehat{t_{f}}]&=&\int_{0}^{\pi} \frac{ds_{1}}{\pi}\int_{0}^{\pi} \frac{ds_{2}}{\pi}\int_{\widehat{t_{i}}}^{\widehat{t_{f}}}\,dt_{1}\int_{\widehat{t_{i}}}^{\widehat{t_{f}}}\,dt_{2}\prod_{n=1}^4\frac{d\epsilon_{n}}{(2\pi)^3}\widehat{J}[\epsilon_{1},s_{1}]\widehat{J}[\epsilon_{2},s_{2}] \nonumber \\ && \delta g_{I} \delta(-\epsilon_{1}+\epsilon_{2}+\epsilon_{3}-\epsilon_{4}) \nonumber \\ &&[[ R^{\dagger (-)}_{\downarrow}(t_{1},-\epsilon_{1};s_{1})(\epsilon_{2}) L^{\dagger (-)}_{\uparrow}(t_{1},\epsilon_{2};s_{1})L^{(-)}_{\uparrow}(t_{2},\epsilon_{3};s_{2})(-\epsilon_{4})R^{(-)}_{\downarrow}(t_{2},-\epsilon_{4};s_{2}) \nonumber \\ & & +L^{\dagger (-)}_{\downarrow}(t_{1},\epsilon_{1};s_{1}) (-\epsilon_{2})R^{\dagger (-)}_{\uparrow}(t_{1},-\epsilon_{2};s_{1})L^{(-)}_{\uparrow}(t_{2},\epsilon_{3};s_{2})(-\epsilon_{4})R{(-)}_{\downarrow}(t_{2},-\epsilon_{4};s_{2})  ]\nonumber \\& &+[ R^{\dagger (-)}_{\downarrow}(t_{1},-\epsilon_{1};s_{1}) (-\epsilon_{1})L^{\dagger (-)}_{\uparrow}(t_{1},\epsilon_{2};s_{1})R^{(-)}_{\uparrow}(t_{2},-\epsilon_{3};s_{2})(\epsilon_{4})L^{(-)}_{\downarrow}(t_{2},\epsilon_{4};s_{2}) +\nonumber \\ & &L^{\dagger (-)}_{\downarrow}(t_{1},\epsilon_{1};s_{1})(\epsilon_{1}) R^{\dagger (-)}_{\uparrow}(t_{1},-\epsilon_{2};s_{1})L^{(-)}_{\uparrow}(t_{2},\epsilon_{3};s_{2})(-\epsilon_{4})R^{(-)}_{\downarrow}(t_{2},-\epsilon_{4};s_{2})  ]]....
\nonumber \\
\end{eqnarray*}

The  spatial derivatives in the last equation  are replaced by the energy  excitations normal to the Fermi Surface $\epsilon$.  As a result,   the engineering dimensions of the coupling constants $\delta g$ is given by: \hspace{0.3 in}
 $\delta g_{I}=\delta\widehat{g}_{I}v_{F}E^{-1}$.
 
The presence of two spatial derivatives  and the two   time integrations generate the engineering  dimensions  $E^{-1}$.  Therefore  we will  ignore  $\delta S^{eff-hole}_{int.}$ for describing the Physics at $low$ $energies$. (The Physics at  $short$ $distances$  is  sensitive  to  operators which  have negative  scaling dimensions   and therefore can not be ignored.)

\end{document}